\documentclass[a4paper, 11pt]{article}

\usepackage{epsf,psfrag}
\catcode`\@=11
\usepackage{cite}
\usepackage{xypic}
\usepackage{indentfirst}
\usepackage{amsmath}
\usepackage{amsfonts}
\usepackage{amssymb}
\usepackage[cp1251]{inputenc}
\usepackage[english]{babel}

\def\II{\hbox{{1}\kern-.25em\hbox{l}}}

\numberwithin{equation}{section}

\begin{document}

\vspace*{1cm}

\begin{center}
{\Large \bf{Separation of variables for the quantum $SL(3,\mathbb{C})$ spin magnet: eigenfunctions of Sklyanin $B$-operator}}

\vspace{0.5cm}

{\large \sf
S. E. Derkachov \footnote{e-mail: derkach@pdmi.ras.ru}, P. A. Valinevich \footnote{e-mail: valinevich@pdmi.ras.ru}}

\vspace{1cm}

{\it St. Petersburg Department of the Steklov Mathematical Institute
of Russian Academy of Sciences,
Fontanka 27, 191023 St. Petersburg, Russia}

\end{center}


\begin{abstract}
The quantum $SL(3, \mathbb{C})$ invariant spin magnet with infinite-dimensional principal series representation in local spaces is considered. We construct eigenfunctions of Sklyanin's $B-$operator which define the representation of separated variables of the model.
\end{abstract}

{\small \tableofcontents}

\newpage

\section{Introduction}

The method of separation of variables has a long-standing history as a tool for treating both classical and quantum-mechanical models. Its essence is the reduction of multidimensional equation to the set of one-dimensional ones.

The quantum separation of variables has been developed by E.K. Sklyanin for the Toda chain in~\cite{SKL1} and for integrable spin chain models
in~\cite{SKL2,SKL3, SKL4,SKL5}. The complete set of SoV states and corresponding scalar products for some models associated with finite-dimensional representations is known explicitly~\cite{Nic, Maillet}. The SOV for the
the quantum integrable models having infinite dimensional
local spaces is considered in~\cite{BT,KM,Sil,KharLeb1,KharLeb2,KharLeb3}.

The main idea of the method is the construction
of the unitary transformation from the initial coordinate representation to the new representation (SOV representation) so that a multi-dimensional
and multi-parameter spectral problem for the transfer matrix in coordinate representation is re-expressed in terms of a multi-parameter but one-dimensional spectral problem. This unitary transformation is realized as integral operator and its kernel is defined by the system of generalized eigenfunctions of some special
Sklyanin operator $B(u)$.
In the simplest case of the symmetry group $SL(2,\mathbb{C})$ operator $B(u)$ coincides with the one of the matrix elements of the monodromy matrix~\cite{SKL2,SKL5}.

For the symmetry group $SL(n,\mathbb{C})$ its construction is more complicated~\cite{SKL3, SKL4,SKL5}.
The classical SOV in this case worked out in~\cite{SKL3,Scott,Gekhtman}
and the algebraic scheme for the quantum SOV in~\cite{SKL4,SKL5,Smirnov}.
It was conjectured in~\cite{Bgood} and proven in~\cite{Slav} that the Sklyanin B-operator can be
used as analog of creation operator
(in closed analogy with algebraic Bethe Ansatz~\cite{KS,Fad1})
for construction of the eigenstates of transfer-matrix in the case of finite-dimensional representations of the symmetry group.

A quantum inverse scattering~\cite{KS,Fad1}
based method for the iterative construction of the generalized eigenfunctions of the B-operator for noncompact $SL(2,\mathbb{C})$-magnet~\cite{FK,L0,L1,L2} is proposed in~\cite{DKM} and then generalized to any matrix element of the monodromy matrix in~\cite{DM}. It results in a pyramidal Gauss-Givental~\cite{GG} representation for the integral kernel of the separation of variables transform.

In the present paper we generalize the iterative construction of the
eigenfunctions of the B-operator for noncompact $SL(3,\mathbb{C})$-magnet.
The whole construction follows the main line of~\cite{DKM,DM} and~\cite{VDKU} where worked out the technique of iterative construction of eigenfunctions of the quantum minors of the $SL(n,\mathbb{C})$-invariant monodromy matrix.
The eigenfunctions are constructed in an explicit form for infinite-dimensional principal series representations of symmetry algebra. This case is simpler than the case of finite-dimensional representations. Namely, there are equivalent representations which differ by the permutation of representation parameters and the corresponding intertwining operators play crucial role in our construction.

The paper is organised as follows. In Sections 2 and 3 we summarize the needed definitions and formulae about $SL(n, \mathbb{C})$ spin magnet and principal series representations of the group $SL(n, \mathbb{C})$. The Section 4 is devoted to the algebraic formulation of the SOV method~\cite{SKL2,SKL3, SKL4,SKL5} for the $SL(2,\mathbb{C})$- and $SL(3,\mathbb{C})$-magnets.
In Section 5 we present the construction of the generalized eigenfunction of the B-operator in the case of inhomogeneous $SL(2,\mathbb{C})$-magnet; this chapter generalizes results of~\cite{DKM,DM} obtained for the homogeneous $SL(2,\mathbb{C})$-magnet.
Construction of the generalized eigenfunction of the B-operator
in the case of inhomogeneous $SL(3,\mathbb{C})$-magnet is
considered in Section 6.

\section{$SL(n,\mathbb{C})$ magnet: description of the model}

In this section we will give the general definition of the quantum magnet associated with $SL(n, \mathbb{C})$ group with special attention to $n=2$ and $n=3$ cases.

The $L-$operator~\cite{Fad1,KS,KRS} for the quantum $SL(n, \mathbb{C})$ magnet
has the form
\begin{equation}
L(u) = u + \sum\limits_{i,j=1}^n e_{ij}E_{ji},\label{ldef}
\end{equation}
where $u$ is the spectral parameter, $e_{ij}$ are the standard matrix units in auxiliary space $\mathbb{C}^n:$ $(e_{ij})_{kl} = \delta_{ik}\delta_{jl},$ and $E_{ij}$ are the generators of $sl(n,\mathbb{C})$ acting on representation space $V.$  They satisfy the standard commutation relations
\begin{equation}
[E_{ij}, E_{kl}] = \delta_{jk}E_{il} - \delta_{il}E_{kj},\quad i,j = 1,\dots, n;\label{slncomm}
\end{equation}
with a constraint
\begin{equation}
E_{11} + E_{22} + \dots + E_{nn} = 0\,.
\end{equation}
The set of the commutation relations can be combined into the well-known Yang-Baxter relation for the product of $L-$operators with different
auxiliary spaces and common quantum space $V:$
\begin{equation}\label{YB}
{\rm R}(u-v) L^{(1)}(u) L^{(2)}(v) = L^{(2)}(v)L^{(1)}(u) {\rm R}(u-v),
\end{equation}
where  ${\rm R}(u)$ is Yang R-matrix ${\rm R}(u)$ defined on $\mathbb{C}^n\otimes \mathbb{C}^n$ as
$$
{\rm R}(u) = u + \sum\limits_{i,j=1}^n e_{ij} \otimes e_{ji}.
$$
For each site $k = 1,\dots,N$ of the magnet one defines a local $L-$operator $L_k(u)$ by (\ref{ldef}) with local $sl(n, \mathbb{C})$ generators $E_{ij}^{(k)},$ $1\le i,j \le n,\ k=1,\dots N.$
The global object for the magnet with $N$ sites is the monodromy matrix
\begin{equation}
T(u) = L_N(u + \delta_N)L_{N-1}(u+ \delta_{N-1})\cdots L_2(u + \delta_2) L_1(u + \delta_1), \label{Top}
\end{equation}
where $\delta_k$ are arbitrary shifts of the spectral parameter. We will consider the general (nonhomogeneous) case $\delta_k \ne 0.$
From (\ref{YB}) and (\ref{Top}) it follows that $T^i_j(u)$ satisfy the set of commutation relations
\begin{equation}
(u-v)[T^i_j(u), T^k_l(v)] = T^k_j(v)T^i_l(u) - T^k_j(u)T^i_l(v),
\label{yangian}
\end{equation}
which define the associative algebra -- Yangian $Y(sl(n, \mathbb{C})).$

The Hilbert space of the model is $V_1 \otimes V_2 \otimes\dots \otimes V_N,$ where $V_k$ is $sl(n, \mathbb{C})$ representation space in $k-$th site. Operators $L_k(u)$ acts nontrivially only on $V_k \otimes \mathbb{C}^n.$ Then $T(u)$ is defined on $V_1 \otimes V_2 \otimes\dots \otimes V_N \otimes \mathbb{C}^n$ and can be presented as a matrix in auxiliary space $\mathbb{C}^n:$
\begin{equation}
T(u) = \left(
\begin{array}{cccc}
T_{\,1}^1(u) & T^1_{\,2}(u) & \cdots & T^1_{\,n}(u)\\
T_{\,1}^2(u) & T^2_{\,2}(u) & \cdots & T^2_{\,n}(u)\\
\vdots & \vdots & \ddots & \vdots\\
T^n_{\,1}(u) & T^n_{\,2}(u) & \cdots & T^n_{\,n}(u)
\end{array}
\right),
\end{equation}
where each entry $T^i_{\,j}(u)$ acts on
$V_1 \otimes V_2 \otimes\dots \otimes V_N.$ We use the similar notations for the entries of $L(u)$ too: $L_k(u)^i_j$ is the element in $i-$th row and $j-$th column of $L_k(u)$, and due to (\ref{ldef}) $L_k(u)^i_j = u\delta_{ij} + E^{(k)}_{ji}.$ Note that leading coefficients in $u$ of $T^i_j(u)$ can be determined by (\ref{Top}):
\begin{equation}
T^i_j(u) = u^N\delta_{ij} + u^{N-1}E_{ji} + O(u^{N-2}),\label{asympt}
\end{equation}
where $E_{ij}$ (without superscripts) are generators of the global $SL(n,\mathbb{C})$ symmetry:
\begin{equation}
E_{ij} = E^{(1)}_{ij}+ E^{(2)}_{ij} + \dots + E^{(N)}_{ij}.\label{globsln}
\end{equation}
We shall use principal series representation of $SL(N,\mathbb{C}),$ which has anti-holomorphic generators $\bar{E}_{ij}$ in addition to holomorphic $E_{ij}.$ They satisfy the same relations as $E_{ij}$ and commute with them.
One can define the anti-holomorphic $L-$operators $\bar{L}_k(\bar{u}) = \bar{u} + \sum\limits_{i,j=1}^n e_{ij}\bar{E}_{ji},$ which depend on the anti-holomorphic spectral parameter $\bar{u},$ and the monodromy matrix $\bar{T}(\bar{u})$ by
\begin{equation}
\bar{T}(\bar{u}) = \bar{L}_N(\bar{u} + \bar{\delta}_N)\bar{L}_{N-1}(\bar{u}+ \bar{\delta}_{N-1})\cdots \bar{L}_2(\bar{u} + \bar{\delta}_2) \bar{L}_1(u + \bar{\delta}_1). \label{Topanti}
\end{equation}
From now on in most part of the paper we will omit formulas concerning anti-holomorphic sector since they are one-to-one with formulas for the holomorphic part.

Commuting hamiltonians of the model are expressed in terms of quantum minors $T^{i_1\dots i_m}_{\,j_1\dots j_m}(u)$ of matrix $T(u)$ (see \cite{MNO,M} for details):
\begin{equation}
T^{i_1\dots i_m}_{\,j_1\dots j_m}(u) = \sum\limits_{\sigma \in S_m} ({\rm sign}\, \sigma)\ T^{i_1}_{\sigma(j_1)}(u) T^{i_2}_{\sigma(j_2)}(u-1)\cdot\dots\cdot T^{i_m}_{\sigma(j_m)}(u-m+1), \label{minor1}
\end{equation}
where $\sigma$ is the permutation of indices $j_1\dots j_m$ and ${\rm sign}\, \sigma$ is its sign.
By the defining relation (\ref{yangian}) of the Yangian, (\ref{minor1}) can be brought to another form
\begin{equation}
T^{i_1\dots i_m}_{\,j_1\dots j_m}(u) = \sum\limits_{\sigma \in S_m} ({\rm sign}\, \sigma)\ T_{j_1}^{\sigma(i_1)}(u-m+1) \cdot\dots\cdot T^{\sigma(i_{m-1})}_{j_{m-1}}(u-1) T_{j_m}^{\sigma(i_m)}(u). \label{minor2}
\end{equation}
Quantum minors are antisymmetric under the permutations of lower and upper indices:
$$
T^{i_1\dots i_s\dots i_r \dots i_m}_{j_1 \dots j_s\dots j_r\dots j_m}(u) = -T^{i_1\dots i_r\dots i_s \dots i_m}_{j_1 \dots j_s\dots j_r\dots j_m}(u) = -T^{i_1\dots i_s\dots i_r \dots i_m}_{j_1 \dots j_r\dots j_s\dots j_m}(u).
$$

Quantum determinant
\begin{equation}\label{qdet}
d(u) = T^{1\dots n}_{\, 1\dots n}(u)
\end{equation}
commutes with all $T^i_j(u)$ and hence is constant on the whole representation space $V_1\otimes\dots \otimes V_N.$

The set of commuting hamiltonians of the $SL(n, \mathbb{C})$ spin magnet is generated by: the set of quantum minors
\begin{equation}
t_k(u) =\sum\limits_{\begin{smallmatrix} i_1,\dots ,i_k=1\\ i_1 < i_2<\dots< i_k  \end{smallmatrix}}^n T^{i_1\dots i_k}_{\,i_1\dots i_k},\quad k=1,\dots, n-1,\label{hams}
\end{equation}
elements $E_{ii},\ i=1,\dots,(n-1)$ of Cartan subalgebra of the global $sl(n, \mathbb{C})$ algebra (\ref{globsln}), and by their antiholomorphic counterparts $\bar{t}_k(\bar{u})$ and $\bar{E}_{kk}.$
In the present paper we will consider only particular cases $n=2$ and $n=3.$

For $n=2$ we will use more customary notations for the generators of the algebra
\begin{equation}\label{lsl2}
L(u) = \left(\begin{array}{cc}
u + E_{11} & E_{21}\\
E_{12} & u+E_{22}
\end{array}\right) = \left(\begin{array}{cc}
u+ S_3 & S_-\\
S_+ & u-S_3
\end{array}\right).
\end{equation}
Generating function of the commuting operators is the transfer-matrix
\begin{equation}
t(u) \equiv t_1(u) = T^{1}_1(u) + T^2_2(u). \label{trans}
\end{equation}
It is polynomial in spectral parameter of degree $N$
$$
t(u) = u^N + \sum\limits_{k=1}^{N-2} t^{(k)} u^k,
$$
and coefficients $t^{(k)}$ commute due to $[t(u), t(v)] = 0.$ To form the complete set of $N$ integrals of motion we add to $(N-1)$ operators $t^{(k)}$ a global operator $S_3 = S_3^{(1)} + S_3^{(2)} + \ldots + S_3^{(N)}.$

For $n=3$ family (\ref{hams}) consists of transfer matrix
\begin{equation}\label{t1}
t_1(u) = T^1_{\,1}(u) + T^2_{\,2}(u) + T_{\,3}^3(u),
\end{equation}
which gives rise to $(N-1)$ integrals of motion, and operator
\begin{equation}\label{t2}
t_2(u) = T^{12}_{\,12}(u) + T_{\,23}^{23}(u) + T^{13}_{\,13}(u),
\end{equation}
which has $(2N-1)$ independent integrals of motion in its decomposition.
%
To complete the set of $3N$ commuting operators we add the global generators $E_{11}$ and $E_{22}.$

\section{Unitary series representations of the $SL(n, \mathbb{C})$ group}

\subsection{General concepts}

In this section we describe the construction of principal series unitary representations of $SL(n, \mathbb{C})$~\cite{GN}. In generic situation these infinite-dimensional representations are irreducible, but at some special values of parameters appears finite-dimensional invariant subspace and the representation  becomes reducible.
In the rest of the paper we will consider only infinite dimensional unitary representations.


Consider two subgroups of $GL(n):$ group $\mathrm{Z}$ of the
lower-triangular complex matrices of the $n$-th order, and the group $\mathrm{H}$ of  upper-triangular complex matrices
\begin{align}\label{zh}
\mathrm{z} = \begin{pmatrix}
1 & 0 & \dots & 0 \\
z_{21} & 1 & \dots & 0 \\
\vdots & \vdots & \ddots & \vdots \\
z_{n1} & z_{n2} & \dots & 1
\end{pmatrix}\in \mathrm{Z}\,,
\qquad
\mathrm{h} = \begin{pmatrix}
h_{11} & h_{12} & \dots & h_{1n} \\
0 & h_{22} & \dots & h_{2n} \\
\vdots & \vdots & \ddots & \vdots \\
0 & 0 & \dots & h_{nn}
\end{pmatrix}\in \mathrm{H}\,.
\end{align}
For almost all $GL(n,\mathbb{C})$ matrices there exists a Gauss decomposition:
matrix $a \in GL(n,\mathbb{C})$
can be presented uniquely as $a =\mathrm{z}\,\mathrm{h}$.

The representation space for operators $T(g)$ is the space of functions $\Phi(\mathrm{z})$, where $\mathrm{z}\in Z$, i.e. $\Phi(\mathrm{z})$ is a function of
$\frac{n(n-1)}{2}$ variables: $\Phi(\mathrm{z})=\Phi(z_{21},
z_{31},\ldots,z_{n,n-1})$. These functions are not assumed to be holomorphic and they also depends on the conjugate variables $\bar z_{21}, \bar
z_{31},\ldots,\bar z_{n,n-1}$. To make formulas more comprehensible we will specify only the holomorphic part of the variables.
Action of operator $T(g)$ on function $\Phi(z)$ is defined by
\begin{align}\label{Tg}
T(g)\,\Phi(\mathrm{z}) = [h_{11}]^{\sigma_1+1}[h_{22}]^{\sigma_2+2}\cdots
[h_{nn}]^{\sigma_n+n}\,
\Phi(\tilde{\mathrm{z}})\,,
\end{align}
where $\mathrm{h}$ and $\tilde{\mathrm{z}}$ are defined by the
Gauss decomposition of the matrix $g^{-1}\mathrm{z} \in GL(n,\mathbb{C}): g^{-1}\mathrm{z} = \tilde{\mathrm{z}}\, \mathrm{h}$.
For the sake of simplicity we will use the compact notation
\begin{equation}\label{[]}
[h]^{\sigma} = h^{\sigma}
\bar{h}^{\bar\sigma}\,,
\end{equation}
where $\bar h$ is complex conjugate of $h$, and complex numbers (representation parameters) $\sigma$ and $\bar\sigma$ differs only by an integer:
$\bar\sigma-\sigma \in \mathbb{Z}$.
The last conditions provides that function $[h]^{\sigma}$ is single-valued.

For the group $SL(n,\mathbb{C})$ we have $\det \mathrm{h}=1,$ and (\ref{Tg}) becomes:
\begin{equation}\label{tsln}
T(g)\,\Phi(\mathrm{z}) =
[\Delta_{1}]^{\sigma_{1}-\sigma_{2}-1}
[\Delta_{2}]^{\sigma_{2}-\sigma_{3}-1}
\cdots[\Delta_{n-1}]^{\sigma_{n-1}-\sigma_{n}-1}\cdot
\Phi(\tilde{\mathrm{z}})\,.
\end{equation}
where $\Delta_k$ is a minor of $\mathrm{h},$ generated by first $k$ rows and columns. As one can see from (\ref{tsln}), representation is purely determined by differences $\sigma_{i}-\sigma_{i+1}$
(together with $\bar\sigma_{i}-\bar\sigma_{i+1}$).
But it is convenient to use the symmetric parametrization
$\boldsymbol{\sigma} =(\sigma_1,\ldots,\sigma_n)$ of the representation
$T^{\boldsymbol{\sigma}}$ of $SL(n,\mathbb{C})$, imposing additional constraint
\begin{equation}\label{rs}
\boldsymbol{\sigma} =(\sigma_1,\ldots,\sigma_n)\ \ ;\ \ \sigma_1+\sigma_2+\ldots+\sigma_n = \frac{n(n-1)}{2}.
\end{equation}
The scalar product in representation space is defined by
\begin{align}\label{scalar}
\langle\Phi_1|\Phi_2\rangle=\int \prod_{1\leq i<k\leq n} d^2 z_{ki}\,
\overline{\Phi_1(\mathrm{z})}\, \Phi_2(\mathrm{z})\,,
\end{align}
where $d^2 z = dxdy$ for $z=x+iy$ and the requirement
of unitarity of the operator $T^{\boldsymbol{\sigma}}(g)$
$$
\langle T^{\boldsymbol{\sigma}}(g)\Phi_1|T^{\boldsymbol{\sigma}}(g)\Phi_2\rangle=\langle \Phi_1|\Phi_2\rangle
$$
leads to the restriction on parameters
$$
\sigma_{k}-\sigma_{k+1} = \bar
\sigma_{k+1}^* -\bar{\sigma}_{k}^*\ \ \ ;\ \ \ k=1,2,\ldots,n-1\,.
$$
Combining it with $\bar{\sigma}_k - \sigma_k \in \mathbb{Z}$ we arrive at the following parametrization for the unitary representations:
\begin{align}\label{Scond}
\sigma_{k}-\sigma_{k+1}=-\frac{n_k}2+i\lambda_k\,, &&
\bar\sigma_{k}-\bar\sigma_{k+1}=\frac{n_k}2+i\lambda_k\,,&&k=1,2,\ldots,n-1\,.
\end{align}
where $n_k \in \mathbb{Z},$ $\lambda_k \in \mathbb{R}.$
Representation $T^{\boldsymbol{\sigma}}$ is irreducible; two representations
$T^{\boldsymbol{\sigma}}$ and $T^{\boldsymbol{\sigma}^{\prime}}$ are
unitary equivalent if and only if~\cite{GN,K} there is a permutation $s$, which transform the set $\sigma_k$ to
$\sigma^{\prime}_k$: $s \boldsymbol{\sigma} = \boldsymbol{\sigma}^{\prime}$.
Intertwining operator $S$
$$
T^{\boldsymbol{\sigma}}\,S = S\,
T^{s \boldsymbol{\sigma}}\,,
$$
realizes unitary equivalence of representations and depends on permutation $s$. All intertwining operators can be constructed from $(n-1)$ basis operators $S_i$ which intertwine representations $T^{\boldsymbol{\sigma}}$ and $T^{\mathrm{s}_i\boldsymbol{\sigma}}$:
$T^{\boldsymbol{\sigma}}\,S_i= S_i\, T^{\mathrm{s}_{i}\boldsymbol{\sigma}}$, where
$\mathrm{s}_{i}\boldsymbol{\sigma}$ differs from $\boldsymbol{\sigma}$ by the transposition of two adjacent parameters:
$$
\mathrm{s}_{i}\,(\ldots\sigma_i,\sigma_{i+1},\ldots)=
(\ldots\sigma_{i+1},\sigma_i,\ldots)
$$
and the same for $\bar\sigma$-parameters:
$\mathrm{s}_{i}\,(\ldots\bar\sigma_i,\bar\sigma_{i+1},\ldots)=
(\ldots\bar\sigma_{i+1},\bar\sigma_i,\ldots)$.
These intertwining operators are well known~\cite{GN,KnS,K} and we shall
use the following convenient explicit expressions for them~\cite{DM3}
\begin{align}\label{Sk}
S_i(\boldsymbol{\sigma})\Phi(\mathrm{z}) =
A\left(\sigma_{i+1}-\sigma_{i}\right)\,
\int d^2 w\,
{[w]^{-1-\sigma_{i+1}+\sigma_{i}}}\,
\Phi\left(\mathrm{z}\,\Bigl(\II-w\,
e_{i+1,i}\Bigr)\right)\,,\\
A(\alpha) =
\frac{i^{\alpha-\bar\alpha}}{\pi}
\,\frac{\Gamma(1+\alpha)}{\Gamma(-\bar\alpha)}\ \ \,, \ \
(e_{ik})_{nm}=\delta_{in}\delta_{km}
\end{align}
Generators of the corresponding Lie algebra can be calculated in a usual way:
for an infinitesimal group element $g = \II + \varepsilon\, e_{ij}$ its representation $T^{\boldsymbol{\sigma}}(g)$ produces the action of the generator $E_{ik}$ on the function of the representation space:
\begin{align}\label{defE}
T(g)\,\Phi(\mathrm{z}) =
\Phi(\mathrm{z})+\left(\varepsilon E_{ik}
+\bar\varepsilon \bar{E}_{ik}\right)
\Phi(\mathrm{z})+O(\varepsilon^2)\,.
\end{align}
They obey the commutation relations (\ref{slncomm}) as far as $e_{ij}$ do.
Generators $E_{ij}$ are differential operators of the first order by construction. Their explicit form for $n=2$ and $n=3$ will be given in next section.

\subsection{Explicit formulae for $n=2,3$}

For the group $\mathrm{GL}(2,\mathbb{C}),$
matrix $\mathrm{z}$ has only one nontrivial entry:
$\mathrm{z}= \left(\begin{smallmatrix} 1&0\\ x& 1 \end{smallmatrix}\right).$
Hence, representation space is space of functions $\Phi(x, \bar{x})$ with
scalar product (\ref{scalar}), i.e. $L_2(\mathbb{C}).$ We omit dependence
on antiholomorphic variables to simplify notations, as it was explained earlier.
%

For the group $SL(2, \mathbb{C})$ we have
\begin{equation}\label{Tsl2}
T^{\sigma_1}\left(g\right)\,\Phi(x)=
\left[d-bx\right]^{2(\sigma_1-1)}\,\Phi\left(\frac{-c+a
x}{d-b x}\right),
\end{equation}
where we have used (\ref{rs}) $\sigma_1 + \sigma_2 = 1.$ Hence its representations are labeled by the only parameter $\sigma_1.$

Generators of the group can be computed by using~(\ref{defE}):
\begin{eqnarray}\label{genssl2}
S_+ = x^2\partial_x - 2(\sigma_1-1)x;\quad S_3 = x\partial_x -(\sigma_1-1);\quad S_- = -\partial_x\,, \label{gensl2}
\end{eqnarray}
where $\partial_x \equiv \frac{\partial}{\partial x}$ and operator $L(u)$ is then
\begin{equation}
L(u) = \left( \begin{array}{cc}
x\partial_x +u_1+1& -\partial_x\\
x\left(x\partial_x +u_1-u_2+1\right) & u_2 - x\partial_x
\end{array}
\right)  = \left(\begin{array}{cc}
1 & 0 \\
x & 1 \end{array}\right) \left(\begin{array}{cc}
u_1 & -\partial_{x} \\
0 & u_2 \end{array}\right) \left(\begin{array}{cc}
1 & 0 \\
-x & 1 \end{array}\right)\label{Laxsl2}
\end{equation}
Note that it is useful to introduce new parameters $u_1 = u-\sigma_1,\ u_2 = u + \sigma_1-1 = u - \sigma_2$ instead of $u$ and $\sigma_1$ and we shall use the following uniform notation for the parameters in L-operator
\begin{align}
L (\mathbf{u}) = L \left(\begin{smallmatrix}
u_{1}\medskip\\
u_{2}
\end{smallmatrix}
\right)\,.
\end{align}
Equivalence of $GL(2, \mathbb{C})$ representations $T^{(\sigma_1, \sigma_2)}$ and $T^{(\sigma_2, \sigma_1)}$ leads to equivalence of $SL(2, \mathbb{C})$ representations $T^{(\sigma_1)}$ and $T^{(1-\sigma_1)}.$ In terms of our new parameters this means that corresponding intertwining operator $S_1=S_1(\mathbf{u})$~(\ref{Sk}) interchanges $u_1$ and $u_2:$
\begin{align}\label{S1defsl2}
L \left(\begin{smallmatrix}
u_{1}\medskip\\
u_{2}
\end{smallmatrix}
\right)\,S_1(\mathbf{u}) = S_1(\mathbf{u})\,L \left(\begin{smallmatrix}
u_{2}\medskip\\
u_{1}
\end{smallmatrix}
\right)\,,\\
\label{S1formsl2}
S_1(\mathbf{u})\Phi(x) = A(u_1-u_2)\,\int d^2 w\,
[w]^{-1-u_1+u_2}\, \Phi(x-w)
\end{align}

For the group $\mathrm{GL}(3,\mathbb{C}),$ matrix $\mathrm{z}$ has three  nontrivial entries and we shall denote the matrix elements
of $\mathrm{z}$ by $x, y, z:$
$
\mathrm{z} = \left( \begin{smallmatrix} 1 & 0 & 0\\ x & 1 &0\\ y & z &1 \end{smallmatrix}\right).
$
Representation is defined on space $L_2(\mathbb{C}^3)$ of functions of three complex variables: $\Phi(\mathrm{z}) = \Phi(x, y, z)$ with scalar product (\ref{scalar}). Explicit form of (\ref{Tg}) becomes rather cumbersome and is not presented here. However, we will extensively use corresponding L-operator and
the following uniform notation for the parameters in L-operator
\begin{align}
L (\mathbf{u}) = L \left(\begin{smallmatrix}
u_{1}\smallskip\\
u_{2}\smallskip\\
u_{3}
\end{smallmatrix}
\right)\ \ ;\ \
u_{1} = u - \sigma_{1}\,,\quad
u_{2} = u - \sigma_{2}\,,\quad
u_{3} = u - \sigma_{3}\,,
\end{align}
where $(\sigma_{1}\,,\sigma_{2}\,,\sigma_3)$ is the set of representation
parameters
\begin{align}
\nonumber
L(\mathbf{u}) =
\left( \footnotesize{ \begin{array}{c|c|c} u_1 + 2 + x\partial_x + y\partial_y & -
\partial_x & -
\partial_y \\ \hline
\begin{array}{c}y\partial_z +\\ x (x\partial_x + y\partial_y - z\partial_z + u_1 - u_2 +1)
\end{array}& u_2 + 1 -x\partial_x + z\partial_z & -\partial_z - x\partial_y\\ \hline
\begin{array}{c} y(x\partial_x +
y\partial_y + z\partial_z+ u_1 - u_3 +2)-\\ -xz(z\partial_z + u_2 - u_3 +1)\end{array}&
\begin{array}{c} - y\partial_x +\\ z(z\partial_z + u_2 - u_3 + 1) \end{array}& u_3 - y\partial_y - z\partial_z
\end{array}}\right)
\label{Laxsl3} = \\
\left(\begin{array}{ccc}
1 & 0 & 0\\
x & 1 & 0\\
y & z & 1
\end{array}\right)
\left(\begin{array}{ccc}
u_1 & -\partial_{x} - z\partial_{y} & -\partial_{y}\\
0 & u_2 & -\partial_{z}\\
0 & 0 & u_3\end{array}\right) \left(\begin{array}{ccc}
1 & 0 & 0\\
-x & 1 & 0\\
xz-y & -z & 1
\end{array}\right)
\end{align}
Intertwining operators $S_1\,,S_2$ satisfy the defining relations
\begin{align}\label{S1sl3}
L \left(\begin{smallmatrix}
u_{1}\smallskip\\
u_{2}\smallskip\\
u_{3}
\end{smallmatrix}
\right)\,S_1(\mathbf{u}) =
S_1(\mathbf{u})\,
L\left(\begin{smallmatrix}
u_{2}\smallskip\\
u_{1}\smallskip\\
u_{3}
\end{smallmatrix}
\right)\ \ ;\ \
L \left(\begin{smallmatrix}
u_{1}\smallskip\\
u_{2}\smallskip\\
u_{3}
\end{smallmatrix}
\right)\,S_2(\mathbf{u}) = S_2(\mathbf{u})\,
L
\left(\begin{smallmatrix}
u_{1}\smallskip\\
u_{3}\smallskip\\
u_{2}
\end{smallmatrix}
\right)
\end{align}
and have the following explicit form
\begin{align}
\label{S12}
S_1(\boldsymbol{u})\Phi(x\,,y\,,z) =
A\left(u_1-u_2\right)\,
\int d^2 w\,
{[w]^{-1-u_1+u_2}}\,
\Phi\left(x-w\,,y-zw\,,z\right)\,,\\
S_2(\boldsymbol{u})\Phi(x\,,y\,,z) =
A\left(u_2-u_3\right)\,
\int d^2 w\,
{[w]^{-1-u_2+u_3}}\,
\Phi\left(x\,,y\,,z-w\right)\,.
\end{align}

\section{SOV for the quantum $SL(2)$ and $SL(3)$ models}

In this section we review the general idea of separation of variables for the quantum integrable systems suggested by E.K. Sklyanin~\cite{SKL1,SKL2,SKL3,SKL4,SKL5} and its application to models related to Yangians $Y[sl(2)]$ and $Y[sl(3)]$~\cite{SKL4}.

\subsection{$SL(2, \mathbb{C})$ magnet}

Consider the magnet with $N$ sites, where in each site $L-$operator is given by (\ref{lsl2}), and spin operators at $k-$th site are realized as differential operators (\ref{genssl2}) with respect to variable $x_k.$ Then the state of the magnet is determined by square-integrable function $\Phi(x) \equiv \Phi(x_1, x_2, \dots, x_N).$

The set of commuting hamiltonians is generated by the coefficients of the holomorphic and anti-holomorphic transfer matrices (\ref{trans}) and operators of the total spin $S_3,$ $\bar{S}_3.$ The problem under consideration is to find their common eigenfunctions and corresponding eigenvalues:
\begin{eqnarray}
t(u)\Phi(x) = \tau(u) \Phi(x);&\quad S_3 \Phi(x) = s_3 \Phi(x),\label{eigtau}\\
\bar{t}(\bar{u})\Phi(x) = \bar{\tau}(\bar{u}) \Phi(x);&\quad \bar{S}_3 \Phi(x) = \bar{s}_3 \Phi(x),
\end{eqnarray}
where $\tau(u), \bar{\tau}(\bar{u})$ are polynomials with complex coefficients, and $s_3, \bar{s}_3 \in \mathbb{C}.$

This problem allows separation of variables if there exists the representation, defined by eigenvalues $q_i$ of some set of operators $\hat{q}_i,\ i=1, \dots, N$ in which the eigenfunction $\Phi(q_1, \dots, q_N)$ factorizes into the product of one-variable functions.
In what follows this representation will be called {\it $q-$representation}, in contrast to the original {\it $x-$representation.}

The construction of such operators $\hat{q}_k$ for models related to the Yangian $Y[SL(2)]$ was presented in \cite{SKL1} and it was adapted to the case of principal series representation of $SL(2, \mathbb{C})$ in \cite{DKM}.

Following these papers, we introduce eigenfunctions $\Psi_p(q|x)$ of operators
\begin{equation}
B(u) = T^1_2(u);\quad \bar{B}(\bar{u}) = \bar{T}^1_2(\bar{u}),
\end{equation}
which are polynomials of the spectral parameter $u$ ($\bar{u}$ respectively) of degree $N-1.$
They are parameterized by roots $q_i$ of their eigenvalues:
\begin{eqnarray}
B(u)\Psi_p(q|x) & = & -ip(u-q_1)\dots(u-q_{N-1})\Psi_p(q|x);\label{Psipdef}\\
\bar{B}(\bar{u})\Psi_p(q|x) & = & -i\bar{p}(\bar{u}-\bar{q}_1)\dots(\bar{u}-\bar{q}_{N-1})\Psi_p(q|x)\label{Psipdef1}.
\end{eqnarray}

The main achievement of \cite{SKL1,DKM} is the proof of the fact that $\Psi_p(q|x)$ performs the transformation to the representation of separated variables, i.e. $\hat{q}_i$ are "operator zeroes" of $B(u):$
$$
B(u) = S_- (u- \hat{q}_1)\dots (u - \hat{q}_{N-1}).
$$
The set of ``coordinates'' in $q-$representation consists of $p, \bar{p},$ and $q = (q_1, \bar{q}_1, \dots q_{N-1}, \bar{q}_{N-1}).$ It will be shown that function $\Psi_p(q|x)$ is well-defined only if they satisfy conditions similar to (\ref{Scond}):
\begin{equation}\label{qcond}
q_k = \frac{m_k}{2} + i\nu_k\,,\quad \bar{q}_k = -\frac{m_k}{2} + i \nu_k,\quad m_k \ \in \mathbb{Z}\ \,,\  \nu_k\ \in \mathbb{R}.
\end{equation}
Note that by (\ref{asympt}) $p$ is the eigenvalue of global generator $S_-:$  $S_-\Psi_p(q|x) = -i p\Psi_p(q|x).$

Here we assume that the spectrum of $B,\bar{B}$ is non-degenerate and $\Psi_p(q|x)$ form a complete orthogonal set on the Hilbert space of the model; there is no exact proof of these statements, but it is strongly believed that they are fulfilled for the wide class of representations. Orthogonality and completeness relations read
\begin{eqnarray}
\int d^{2N}x\, \Psi_p(q|x) \overline{\Psi}_{p'}(q'|x) =
\mu^{-1}(p, q)\,\delta^2(\vec{p}-\vec{p}^{\,\prime})\,
\delta_{N-1}(q-q');\label{Psiort}\\
\int d^{2}p\,\int \mathcal{D}_{N-1}q \, \mu(p, q)\, \Psi_p(q|x) \overline{\Psi}_p(q|x')  =  \prod_{i=1}^{N}\delta^2(\vec{x}_i - \vec{x}_i^{\,\prime}),\label{Psicompl}
\end{eqnarray}
where delta-functions and integration measures should be understood as follows. In $x-$representation integration with measure $d^{2N}x = \prod\limits_{i=1}^N d^2x_i$ is by definition the scalar product~(\ref{scalar}), and $\delta^2(\vec{x} - \vec{x}') = \delta(x-x')\delta(\bar{x} - \bar{x}').$ Integration in $q-$representation due to (\ref{qcond}) is understood as
$$
\int \mathcal{D}_{N-1}q  =  \prod\limits_{k=1}^{N-1}
\left(\sum\limits_{m_k = -\infty}^{\infty}\int\limits_{-\infty}^{\infty} d\nu_k\right)
$$
and delta-function in $q-$representation is defined as
symmetrized expression
\begin{align}\label{symdelt}
\delta_N(q-q')=\frac1{N!}
\sum_{s\in S_N}\prod_{k=1}^N\delta^{(2)}(q_k-q'_{s(k)}),
\end{align}
where the sum goes over all permutations of $N$ elements, and
\begin{align}
\delta^{(2)}(q-q')\equiv
\delta_{m m'} \delta(\nu-\nu').
\end{align}
Integration over $\vec{p} = (p_1,p_2),$ where $p = p_1+ip_2\,,\bar{p} = p_1-ip_2,$ goes over the whole complex plane, and $d^2p = dp_1 dp_2;$ corresponding delta-function $\delta^2(\vec{p} - \vec{p}') = \delta(p-p')\delta(\bar{p} - \bar{p}').$

In this notations transition from one representation to another has the following form:
\begin{eqnarray}
\Phi(x) & = & \int d^{2}p\,\int \mathcal{D}_{N-1}q\, \mu(p, q)\, \Psi_p(q|x) \Phi(p,q),\label{xtopq}\\
\Phi(p,q) & = & \int d^{2N}x\, \overline{\Psi}_p(q|x) \Phi(x),\label{pqtox}
\end{eqnarray}
The weight function $\mu(p, q)$ is the Sklyanin measure and
here we do not need its explicit form~\cite{DKM}.
To prove the fact that in $q-$representation the eigenfunction of hamiltonians $\Phi$ factorizes
\begin{equation}\label{factor}
\bar{\Phi}(p, q) = \phi_0(p)\phi_1(q_1)\dots \phi_{N-1}(q_{N-1}),
\end{equation}
we need the following three relations connecting $B(u)$ and operator
$A(u) = T^1_2(u).$

{\bf Proposition 1}
\begin{eqnarray}
[B(u), B(v)] = 0\ \,, \label{prop1}\\
\label{prop2}
\medskip
(u-v+1) A(u) B(v) = (u-v) B(v) A(u) + A(v) B(u)\ \,,\\
\label{prop3}
\medskip
A(u+1)A(u) - t(u+1) A(u) + d(u+1) = -T^2_1(u+1) B(u)\ \,.
\end{eqnarray}
Similar formulas hold for antiholomorphic operators as well. Relations~(\ref{prop1}) and~(\ref{prop2}) are particular cases of commutation relations (\ref{yangian}) and (\ref{prop3}) follows from one of the forms of quantum determinant (\ref{qdet}):
$$
d(u) = T_2^2(u)T_1^1(u-1) - T^2_1(u) T^1_2(u-1),
$$
which is a constant on the whole representation space.

Due to (\ref{prop1}) coefficients $b_k$ in decomposition of $B(u),$ $B(u) = \sum b_k u^k$ form a commuting family of operators, $[b_k, b_m] = 0.$ Applying the rhs and lhs of (\ref{prop2}) to the function $\Psi_p(q|x)$ and taking $u = q_i,$ we get
\begin{equation}
B(v)A(q_i)\Psi_p(q|x) = -ip(v-q_1)\cdots(v-q_i-1)\cdots(v - q_{N-1})A(q_i)\Psi_p(q|x).\label{APsi}
\end{equation}
Moreover, $\bar{B}(\bar{v}) A(u) = A(u) \bar{B}(\bar{v}),$ and hence
\begin{equation}
\bar{B}(\bar{v}) A(q_i) \Psi_p(q|x) = -i\bar{p}(\bar{v} - \bar{q}_1)\cdots(\bar{v} - \bar{q}_i)\cdots(\bar{v} - \bar{q}_{N-1}) A(q_i) \Psi_p(q|x).\label{barAPsi}
\end{equation}
From (\ref{APsi}) and (\ref{barAPsi}) follows that $A(q_i)\Psi_p(q|x)$ is proportional to $\Psi_p(E_i^+q|x),$  where $E_i^+q$ stands for the set $q$ with element $q_i$ shifted by +1 and other elements unchanged: i.e if
$$
q = (q_1, \dots, q_i, \dots, q_N; \bar{q}_1, \dots, \bar{q}_N),
$$
then
$$
E_i^+ q = (q_1, \dots, q_i+1, \dots, q_N; \bar{q}_1, \dots, \bar{q}_N).
$$

In analogous way one can show that $ \bar{A}(\bar{q}_i)\Psi_p(q|x) \sim \Psi_p(\bar{E}_i^+q|x),$
where the set
$$
\bar{E}_i^+ q = (q_1, \dots, q_N; \bar{q}_1, \dots, \bar{q}_i+1, \dots, \bar{q}_N).
$$
By the choice of normalization for $\Psi$ we can make proportionality coefficient equal to unity, i.e.
\begin{equation}\label{Aaction}
A(q_i)\Psi_p(q|x) = \Psi_p(E_i^+q|x),\quad \bar{A}(q_i) \Psi_p(q|x) = \Psi_p(\bar{E}_i^+ q|x).
\end{equation}

Now let us show that (\ref{prop3}) leads to separated equations for the function $\Psi.$ Applying both sides of it to $\Psi_p(q|x)$ and putting $u = q_i$ with the help of (\ref{Aaction}) we get
\begin{equation}\label{presep}
\Psi_p(E_i^{+2}q|x) - t(q_i+1) \Psi_p(E_i^+q|x) + d(q_i+1)\Psi_p(q|x) = 0
\end{equation}
(rhs is zero due to (\ref{Psipdef})). Multiplying (\ref{presep}) by $\bar{\Phi}(x)$ and integrating over $x,$ with the use of (\ref{pqtox}) and (\ref{eigtau}) we obtain equation for conjugated function $\bar{\Phi}(p, q):$
\begin{equation}
\bar{\Phi}(p,E_i^{+2}q) - \tau(q_k+1)\bar{\Phi}(p,E_k^+q) + d(q_k+1)\bar{\Phi}(p, q) = 0, \quad \forall k = 1, \dots, N-1,\label{sepsl2}
\end{equation}
where $\tau(q_i+1)$ is the eigenvalue of the transfer-matrix $t(u)$ at $u = q_i+1.$ This set of equations can be solved by the ansatz $\bar{\Phi}(p, q) = \phi_0(p)\phi_1(q_1)\dots \phi_{N-1}(q_{N-1}),$ which leads to separated equations
\begin{equation}
\phi_k(q_k+2) - \tau(q_k+1)\phi_k(q_k+1) + d(q_k+1)\phi_k(q_k) = 0.\label{Sepsl2}
\end{equation}
Now let us write the equation for the function $\phi_0(p).$ Expanding (\ref{prop2}) in powers of $u$ one gets
\begin{equation}
[S_3, B(v)] = - B(v),
\end{equation}
from which it follows that
\begin{equation}
\lambda^{S_3} B(u) \lambda^{-S_3} = \lambda^{-1} B(u),
\end{equation}
where  $\lambda \in \mathbb{C},$ and $\lambda^{S_3}$ is understood as the formal power series in $\lambda.$ Applying the last operator equation to the function $\lambda^{S_3} \Psi_p(q|x)$ we see that $\lambda^{S_3} \Psi_p(q|x) \sim \Psi_{\lambda p}(q|x),$ i.e. global transformations of the form $\lambda^{S_3}$ generate the scaling of the parameter $p.$ From the fact that these transformations form a one parameter group, one derives that
\begin{equation}
\lambda^{S_3} \Psi_p(q|x) = \lambda^m \Psi_{\lambda p}(q|x),
\end{equation}
where parameter $m$ is determined by $p-$dependent part of the normalization coefficient of $\Psi_{p}(q|x).$ The change of normalization $\Psi_{p}(q|x) \to p^k \Psi_{ p}(q|x)$ changes $m \to m+k$ and measure $\mu(p, q) \to [p]^{-k} \mu(p,q),$ but does not affect other properties of eigenfunction including (\ref{Aaction}).

For infinitesimal transformations $\lambda = 1 + \epsilon$ one has
\begin{equation}
S_3 \Psi_p(q|x) = (p \partial_p + m)\Psi_p(q|x).
\end{equation}
It allows us to derive the equation for the function $\phi_0(p):$
\begin{equation}
s_3 \phi_0(p) = (p \partial_p + m) \phi_0(p).
\end{equation}

\subsection{$SL(3,\mathbb{C})$ magnet}

Similar to $SL(2, \mathbb{C})$ case, operator that performs the transfer to representation of separated variables ($q-$representation) is generated by eigenfunctions of the certain operator $B(u),$ and there exists an operator $A(u)$ that acts as a shift operator on these functions (see (\ref{Aaction})). But expressions for $A(u)$ and $B(u)$ now are more involved. Namely,
\begin{equation}
B(u) = T^2_3(u) T_{23}^{12}(u+1) + T_3^1(u) T_{13}^{12}(u+1), \label{Bsl3}
\end{equation}
\begin{equation}
A(u) =  T_{12}^{13}(u+1) \left( T_3^2(u+1) \right)^{-1}, \label{Asl3}
\end{equation}
We deal with the complex algebra $sl(3, \mathbb{C})$ and, as usual, define anti-holomorphic operators $\bar{B}(\bar{u})$ and $\bar{A}(\bar{u})$ in terms of $\bar{T}^i_j(\bar{u})$ by (\ref{Bsl3})--(\ref{Asl3}).

Using defining relations of the Yangian together with commutators of
matrix elements and minors,
it was shown in \cite{SKL4} that $A(u)$ and $B(u)$ have properties analogous to (\ref{prop1})-(\ref{prop3}) of previous section.

{\bf Proposition 2.}
\begin{align}\label{commsl3}
[B(u), B(v)] = 0\ \ \,,\ \ \quad [A(u), A(v)] = 0\ \,,
\end{align}
\begin{align}
\label{ABsl3}
(u+v-1)A(u)B(v) - (u-v)B(v)A(u)= \\
\nonumber
=\left( T^2_3(u+1) \right)^{-1} T_3^2(v) \left( T_3^2(u) \right)^{-1} T^2_3(v+1)A(v)B(u)\ \,,
\end{align}
\begin{align}
\label{BTsl3}
A(u+2)A(u+1)A(u) + t_1(u+2)A(u+1)A(u) + t_2(u+2)A(u) + d(u+2) = \\
\nonumber
= \left( T^2_3(u+2)T^{23}_{13}(u+2) - T^2_1(u+2)T^{12}_{13}(u+2) \right)\left( T^2_3(u)T^2_3(u+1) T^2_3(u+2) \right)^{-1} B(u)\ \,.
\end{align}

The last equation involves two transfer matrices $t_1(u)$ and $t_2(u)$ given by (\ref{t1}), (\ref{t2}), and quantum determinant $d(u)$. The proof of (\ref{commsl3})-(\ref{BTsl3}) is much more complicated in comparison to $SL(2, \mathbb{C})$ case. We refer reader to the original paper \cite{SKL4} for details\footnote{Equations (\ref{ABsl3}) and (\ref{BTsl3}) differs from those presented in \cite{SKL4} since we use different rule for defining operator-valued finctions. See \cite{SKL4} for details.}.

It follows from (\ref{Bsl3}) that $B(u)$ is the polynomial on $u$ of degree $3N-3$ and its eigenfunctions are characterized by $3N-2$ parameters $p,\ q_i:$
$$
B(u)\Psi = ip\prod\limits_{i=1}^{3N-3} (u-q_i) \Psi.
$$
The same holds for its anti-holomorphic counterpart $\bar{B}(\bar{u}):$ $ \bar{B}(\bar{u})\Psi = i\bar{p}\prod_{i=1}^{3N-3} (\bar{u}-\bar{q}_i) \Psi$ with anti-holomorphic parameters $\bar{p},\ \bar{q}_i.$

Coefficients of $B(u),\ \bar{B}(\bar{u})$ form a commutative family of $6N-4$ operators, while the quantum system has $6N$ degrees of freedom. To have the complete set of commuting operators, we have to add two more operators both in holomorphic and antiholomorphic sector. We choose them to be generators of global $sl(3,\mathbb{C})$ algebra $E_{32},\ E_{31}$ (see (\ref{globsln})) and $\bar{E}_{32},\ \bar{E}_{31}.$ One can check that they commute with $B(u),\ \bar{B}(\bar{u})$ and among themselves.
So, variables in $x-$representation are $x_i, y_i, z_i$ $(i=1, \dots, N),$ or, in short $x, y, z,$ where $x = (x_1, \dots, x_N)$ etc. together with their antiholomorphic counterparts $\bar{x}_i, \bar{y}_i, \bar{z}_i$. In $q-$representation variables are $p_1, p_2, p, q,$ where $q = (q_1, \dots, q_{3N-3})$ and their antiholomorphic counterparts ($p_1$ and $p_2$ are eigenvalues of global generators $E_{32},\ E_{31}$). All eigenfunction will be well-defined if $q_i$ are of the form (\ref{qcond}).
Eigenfunctions under consideration will be denoted by $\Psi_{p_1p_2p}(q|x, y, z).$ They satisfy the set of equations
\begin{eqnarray}
B(u)\Psi_{p_1p_2p}(q|x, y, z) &\!\!=\!\!& ip\prod\limits_{i=1}^{3N-3} (u-q_i) \Psi_{p_1p_2p}(q|x, y, z)\nonumber\\
E_{31}\Psi_{p_1p_2p}(q|x, y, z) &\!\!=\!\!&-ip_1 \Psi_{p_1p_2p}(q|x, y, z) \label{Psisl3} \\
E_{32} \Psi_{p_1p_2p}(q|x, y, z) &\!\!=\!\!& -ip_2 \Psi_{p_1p_2p}(q|x, y, z),\nonumber
\end{eqnarray}
and similar equations for the antiholomorpic part.
Integral transformations that bring the function $\Phi$ from $x-$ to $q-$representation and back read (cf. (\ref{xtopq}) and (\ref{pqtox})):
\begin{equation}
\Phi(p_1, p_2, p, q) = \int d^{2N}x\, d^{2N}y\, d^{2N}z\,\Phi(x,y,z)\overline{\Psi}_{p_1p_2 p}(q|x, y, z);\label{Phixsl3}
\end{equation}
\begin{equation}
\Phi(x, y, z) = \int d^2p_1\, d^2p_2\, d^2p\,\int \mathcal{D}_{3N-3}q\, \mu(p_1, p_2, p,q)\,\Phi(p_1, p_2, p,q) \Psi_{p_1p_2 p}(q|x, y, z)\,,\label{Phizsl3}
\end{equation}
where the integration in $q-$representation due to (\ref{qcond})
is understood as
$$
\int \mathcal{D}_{3N-3}q  =  \prod\limits_{k=1}^{3N-3}
\left(\sum\limits_{m_k = -\infty}^{\infty}\int\limits_{-\infty}^{\infty} d\nu_k\right)
$$
Function $\mu(p_1, p_2, p, q)$ originates from the normalization condition for eigenfunctions:
\begin{multline}
\int d^{2N}x\, d^{2N}y\, d^{2N}z\, \overline{\Psi_{p'_1p'_2 p'}(q'|x, y, z)} \Psi_{p_1p_2p}(q|x, y, z) =\\= \mu^{-1}(p_1,p_2,p,q)\,\delta^{2}(\vec{p}_1 - \vec{p}_1^{\,\prime})\,\delta^{2}(\vec{p}_2 - \vec{p}_2^{\,\prime})\,\delta^{2}(\vec{p}-\vec{p}^{\,\prime})\,\delta_{3N-3}(q-q').
\end{multline}
with symmetrized delta function $\delta_{3N-3}(q-q')$ of the form (\ref{symdelt}).

In full analogy to $SL(2, \mathbb{C})$ case, (\ref{ABsl3}) implies that operator $A(u)$ is a shift operator for $\Psi_{p_1p_2p}(q|x, y, z):$
\begin{equation}
A(q_i)\Psi_{p_1p_2p}(q|x, y, z) = \Psi_{p_1p_2p}(E_i^+q|x, y, z),\quad i=1,\dots,3N-3.
\end{equation}
Now consider $\Phi$ -- eigenfunction of hamiltonians:
\begin{equation}
t_1(u)\Phi = \tau_1(u)\Phi;\quad t_2(u)\Phi = \tau_2(u)\Phi.
\end{equation}
Applying both sides of (\ref{BTsl3}) to $\Psi_{p_1p_2p}(q|x, y, z),$ integrating with $\overline{\Phi}(x, y, z)$ over $x, y, z$ we arrive at the equation for $\Phi$ in $q-$representation
\begin{multline}
\bar{\Phi}(p_1, p_2, p, E_i^{+3}q) + t_1(q_i+2) \bar{\Phi}(p_1, p_2, p, E_i^{+2}q) + \\
+t_2(q_i+2)\bar{\Phi}(p_1, p_2, p, E_i^+q) + d(q_i+2)\bar{\Phi}(p_1, p_2, p, q) = 0\,.\label{sepsl3}
\end{multline}
It can be reduced to the set of one-dimensional equations
\begin{equation}
\varphi(q_i+3) + t_1(q_i+2)\varphi(q_i+2) + t_2(q_i+2)\varphi(q_i+1) + d(q_i+2)\varphi(q_i) = 0,
\label{Sepsl3}
\end{equation}
if we take the ansatz
$$
\bar{\Phi}(p_1, p_2, p, q) = \varphi_{11}(p_1) \varphi_{22}(p_2) \varphi_0(p)\varphi_1(q_1)\dots \varphi_{3N-3}(q_{3N-3}).
$$
Separated equations for $\phi_{11}(p_1),\ \phi_{22}(p_2)$ follow from
\begin{equation}
\lambda^{E_{11}}E_{31}\lambda^{-E_{11}} = \lambda^{-1}E_{31};\quad \lambda^{E_{22}}E_{32}\lambda^{-E_{22}} = \lambda^{-1}E_{32}
\end{equation}
and read
\begin{eqnarray}
e_{11}\,\phi_{11}(p_1) &= \left(p_1\partial_{p_1} + m_1\right)\phi_{11}(p_1);\\
e_{22}\,\phi_{22}(p_2) &= \left(p_2\partial_{p_2} + m_2\right)\phi_{22}(p_2),
\end{eqnarray}
where $e_{11},\ e_{22}$ are values of the integrals of motion $E_{11},\ E_{22}$ on $\Phi.$ Parameters $m_1, m_2$ are arbitrary and are defined by the $p_1$- and $p_2$-dependent part of normalization coefficient of $\Psi_{p_1p_2p}(q|x, y, z).$

\section{Eigenfunctions of the operator $B(u)$ for the $SL(2, \mathbb{C})$ magnet.}

\subsection{Permutation of parameters in the monodromy matrix}

In this section we will construct the manifest form of functions $\Psi_p(q|x)$ satisfying (\ref{Psipdef})-(\ref{Psipdef1}). Our construction heavily relies on the existence of the intertwining operator (\ref{S1defsl2})-(\ref{S1formsl2})
which interchanges parameters $u_1$ and $u_2$ inside L-operator
\begin{align*}
L \left(\begin{smallmatrix}
u_{1}\medskip\\
u_{2}
\end{smallmatrix}
\right)\, S_1(\mathbf{u}) = S_1(\mathbf{u})\,L \left(\begin{smallmatrix}
u_{2}\medskip\\
u_{1}
\end{smallmatrix}
\right)
\end{align*}
and an operator $S$ which interchanges parameters $u_1\rightleftarrows v_2$ in the product of two $L-$operators:
\begin{equation}
L_2\left(\begin{smallmatrix}
u_{1}\medskip\\
u_{2}
\end{smallmatrix}
\right) L_1 \left(\begin{smallmatrix}
v_{1}\medskip\\
v_{2}
\end{smallmatrix}
\right)\,S = S\,L_2 \left(\begin{smallmatrix}
v_{2}\medskip\\
u_{2}
\end{smallmatrix}
\right) L_1 \left(\begin{smallmatrix}
v_{1}\medskip\\
u_{1}
\end{smallmatrix}
\right)\,
\label{S2def}
\end{equation}
In this formula $L-$operators have the same auxiliary space and different quantum spaces, i.e. matrix elements of $L_2(u_1, u_2)$ and $L_1(v_1, v_2)$ are differential operators with respect to variables $x_2$ and $x_1$ correspondingly.
It can be shown~\cite{DM3} that $S$ is the operator of multiplication by the simple function
\begin{equation}
S(u_1 - v_2) = [x_2 - x_1]^{v_2-u_1}\,
\end{equation}
We shall use the following uniform notation for the parameters in L-operator for the k-th site
\begin{align}
L_k(\mathbf{u}_k) = L_k\left(\begin{smallmatrix}
u_{1 k}\medskip\\
u_{2 k}
\end{smallmatrix}
\right)\ \ ;\ \
u_{1\,k} = u - \sigma_{1}^{(k)} + \delta_k;\quad
u_{2\,k} = u - \sigma_{2}^{(k)} + \delta_k\,,
\end{align}
where $(\sigma_{1}^{(k)}\,,\sigma_{2}^{(k)})$ is the set of representation parameters in the k-th site.
Elements of the monodromy matrix from site k to site n
\begin{equation}\label{mon}
T(U) = L_n(\mathbf{u}_n)L_{n-1}(\mathbf{u}_{n-1})
\cdots L_{k+1}(\mathbf{u}_{k+1}) L_k(\mathbf{u}_k),
\end{equation}
depend on the set of parameters $\mathbf{u}_{i}$, where $k\leq i \leq n$ and we combine all parameters in the matrix $U$
\begin{equation}
U = \left(
\begin{array}{ccccc}
u_{1\,n} & u_{1\,n-1} & \dots & u_{1\,k+1} & u_{1\,k} \\
u_{2\,n} & u_{2\,n-1} & \dots & u_{2\,k+1} & u_{2\,k}
\end{array}
\right)\,, \label{parsl2}
\end{equation}
where the $i-$th column of this matrix contains parameters of the i-th
L-operator $L_i(\mathbf{u}_i)$.

Let us introduce the intertwining operators $S_1$ for each site of the chain.
We will denote them $S_{1}(\mathbf{u}_k)$ and each of them interchanges parameters
$u_{1 k}\rightleftarrows u_{2 k}$ inside the L-operator at k-th site
\begin{equation}
L_k\left(\begin{smallmatrix}
u_{1 k} \medskip \\
u_{2 k}
\end{smallmatrix}
\right)\, S_1(\mathbf{u}_k) = S_1(\mathbf{u}_k)\,L_k\left(\begin{smallmatrix}
u_{2 k}\medskip\\
u_{1 k}
\end{smallmatrix}
\right)\,
\label{S1k}
\end{equation}
Next we introduce the operators $S$ for each pair of two adjacent cites.
We will denote them $S(\mathbf{u}_{k+1}\,,\mathbf{u}_k)$ and each of them interchanges parameters
$u_{1\,k+1}\rightleftarrows u_{2\,k}$ inside the product of L-operators
at two adjacent sites
\begin{gather}
T\left(\begin{smallmatrix}
u_{1\,k+1} & u_{1\,k} \medskip\\
u_{2\,k+1} & u_{2\,k}
\end{smallmatrix}
\right) = L_{k+1}(\mathbf{u}_{k+1}) L_k(\mathbf{u}_k)\, ; \\
T\left(\begin{smallmatrix}
u_{1\,k+1} & u_{1\,k} \medskip\\
u_{2\,k+1} & u_{2\,k}
\end{smallmatrix}
\right)\,
S(\mathbf{u}_{k+1}\,,\mathbf{u}_k) = S(\mathbf{u}_{k+1}\,,\mathbf{u}_k)\,T\left(\begin{smallmatrix}
u_{2\,k} & u_{1\,k} \medskip\\
u_{2\,k+1} & u_{1\,k+1}
\end{smallmatrix}
\right)
\label{S2k}
\end{gather}
The explicit formulae for these elementary intertwining operators are
\begin{align}\label{S1Sk}
S_1(\mathbf{u}_k)\Phi(x_k) = S_1(u_{1k}-u_{2k})\Phi(x_k) = A(u_{1k}-u_{2k})\,\int d^2 w\,
[w]^{-1-u_{1k}+u_{2k}}\, \Phi(x_k-w)\, ; \\
S(\mathbf{u}_{k+1}\,,\mathbf{u}_k) = S(u_{2k}-u_{1k+1}) =
[x_{k+1} - x_{k}]^{u_{2k}-u_{1k+1}}
\end{align}
Clearly, $[S_1(\mathbf{u}_k)\,,L_i(\mathbf{u}_i)] = 0 $ for $i \ne k$
and $[S(\mathbf{u}_{k+1}\,,\mathbf{u}_k)\,,L_{i}(\mathbf{u}_{i})] = 0$ for $i \ne k, k+1$ so that for the complete monodromy matrix
from the first site to the N-th site we obtain
\begin{gather}
T\!\left(\begin{smallmatrix}
u_{1N} & \dots & u_{1k} & \dots & u_{1 1}\medskip\\
u_{2N} & \dots & u_{2k} & \dots & u_{2 1}
\end{smallmatrix}
\right)\,S_1(\mathbf{u}_k) = S_1(\mathbf{u}_k)\,T\!\left(\begin{smallmatrix}
u_{1N} & \dots & u_{2k} & \dots & u_{11} \medskip \\
u_{2N} & \dots & u_{1k} & \dots & u_{21}
\end{smallmatrix}
\right)\,, \\
T\left(\begin{smallmatrix}
u_{1 N} & \dots & u_{1\,k+1} & u_{1\,k} & \dots & u_{11} \medskip\\
u_{2 N} & \dots & u_{2\,k+1} & u_{2\,k} & \dots & u_{21}
\end{smallmatrix}
\right)\, S(\mathbf{u}_{k+1}\,,\mathbf{u}_k) =
S(\mathbf{u}_{k+1}\,,\mathbf{u}_k)\,T\left(\begin{smallmatrix}
u_{1 N} & \dots & u_{2\,k} & u_{1\,k} & \dots & u_{1 1} \medskip\\
u_{2 N} & \dots & u_{2\,k+1} & u_{1\,k+1} & \dots & u_{2 1}
\end{smallmatrix}
\right)\,.
\end{gather}
Hence, taking products of $S_1(\mathbf{u}_k)$ and $S(\mathbf{u}_{k+1}\,,\mathbf{u}_k)$ with suitable arguments, we can construct operator which performs {\it any}  permutation of elements in $U.$

\subsection{Change of parameters in operator $B(u)$}

Let us find out parameter dependence of $B(u) = T^1_2(u).$ Being matrix element of $T(U),$ $B(u) = B(U)$ with $U$ defined for the complete monodromy matrix
\begin{equation}
U = \left(
\begin{array}{cccc}
u_{1 N} & \dots & u_{1 2} & u_{1 1}\\
u_{2 N} & \dots & u_{2 2} & u_{2 1}
\end{array}
\right) \label{parsl2}
\end{equation}
From (\ref{Laxsl2}) we see that first line of $L(u)$ does not contain $u_2.$ By (\ref{Top})
\begin{equation}
B(u) = \sum\limits_{a,b\ldots, c =1\,,2}
L_N(u)^1_{a}\, L_{N-1}(u)^{a}_{b}
\cdots L_1(u)^{c}_2,
\end{equation}
hence $u_{2N}$ is not present in $B(u)$ and we have
\begin{align}\label{v}
B\left(\begin{smallmatrix}
u_{1N} & \dots & u_{12} & u_{11}\medskip\\
u_{2N} & \dots & u_{22} & u_{21}
\end{smallmatrix}
\right) = B \left(\begin{smallmatrix}
u_{1N} & \dots & u_{12} & u_{11}\medskip\\
v & \dots & u_{22} & u_{21}
\end{smallmatrix}
\right)
\end{align}
with arbitrary new parameter $v$. This parameter can be transferred to any position in $U$ with the use of intertwining operators (\ref{S1k}), (\ref{S2k}). It can be demonstrated by the following sequence of transformations:
\begin{multline}\label{idea}
B\! \left(\begin{smallmatrix}
u_{1N} & \dots & u_{12} & u_{11}\medskip\\
u_{2N} & \dots & u_{22} & u_{21}
\end{smallmatrix}
\right)\, S_1(u_{1N}-v)  = B\! \left(\begin{smallmatrix}
u_{1N} & \dots & u_{12} & u_{11}\medskip\\
v & \dots & u_{22} & u_{21}
\end{smallmatrix}
\right)\, S_1(u_{1N}-v) =\\
= S_1(u_{1N}-v)\, B\! \left(\begin{smallmatrix}
v & \dots & u_{12} & u_{11}\medskip\\
u_{1N} & \dots & u_{22} & u_{21}
\end{smallmatrix}
\right) = S_1(u_{1N}-v)\, B\! \left(\begin{smallmatrix}
v & \dots & u_{12} & u_{11}\medskip\\
u_{2N} & \dots & u_{22} & u_{21}
\end{smallmatrix}
\right).
\end{multline}
As a result, we get $B\left(\begin{smallmatrix}
v & \dots & u_{12} & u_{11}\medskip\\
u_{2N} & \dots & u_{22} & u_{21}
\end{smallmatrix}
\right)$ with parameter matrix which element $(1N)$ is now arbitrary. This idea can be used to substitute {\it any} element (and any number of elements) of $U$ by arbitrary parameter(s).

\subsection{Eigenfunctions of $B(u)$}
\label{Bsl2}

Consider operator $W(U, V)$ which intertwines $B(U)$ with $B(V)$
\begin{align}\label{BW}
B(U)\, W(U, V) = W(U, V)\, B(V)\,,
\end{align}
where
\begin{align}
U  = \left(
\begin{array}{ccccc}
u_{1 N} & u_{1\, N-1} & \dots & u_{1 2} & u_{1 1}\\
u_{2 N} & u_{2\, N-1} & \dots & u_{2 2} & u_{2 1}
\end{array}
\right)\ \ ;\ \ V  = \left(
\begin{array}{ccccc}
u_{1 N} & u_{1\, N-1} & \dots & u_{1 2} & u_{1 1}\\
u_{2 1} & v_{N-1} & \dots & v_{2} & v_{1}
\end{array}
\right)
\end{align}
Note that due to~(\ref{v}) there is not any dependence on the
parameters $u_{2N}$ and $u_{21}$, so that the operator $W(U, V)$ effectively contains $\textstyle(N-1)$ arbitrary parameters $v_{1},v_{2},\ldots,v_{N-1}$ and it can be constructed from the elementary intertwining operators in a many equivalent ways.
We give a more or less canonical construction using operators
$R_{k+1 k}$~\cite{VDKU,DM1,DM3}
each of them interchanges parameters
$u_{2\,k+1}\rightleftarrows u_{2\,k}$ at two adjacent sites inside the product of L-operators
\begin{gather}\nonumber
T\left(\begin{smallmatrix}
\dots & u_{1\,k+1} & u_{1\,k} & \dots \medskip\\
\dots & u_{2\,k+1} & u_{2\,k} & \dots
\end{smallmatrix}
\right)\, R_{k+1 k}\left(\begin{smallmatrix}
u_{1\,k+1} & u_{1\,k} \medskip\\
u_{2\,k} & u_{2\,k+1}
\end{smallmatrix}
\right) = R_{k+1 k}\left(\begin{smallmatrix}
u_{1\,k+1} & u_{1\,k} \medskip\\
u_{2\,k} & u_{2\,k+1}
\end{smallmatrix}
\right)\,\, T\left(\begin{smallmatrix}
\dots & u_{1\,k+1} & u_{1\,k} & \dots \medskip\\
\dots & u_{2\,k} & u_{2\,k+1} & \dots
\end{smallmatrix}
\right)
\end{gather}
Note that the parameters in R-matrix mimic exactly parameters in
the monodromy matrix in the right hand side of the considered relation.
The chain of the elementary transpositions
\begin{align}\nonumber
\left(\begin{smallmatrix}
u_{1\,k+1} & u_{1\,k} \medskip\\
u_{2\,k} & u_{2\,k+1}
\end{smallmatrix}
\right)\xleftarrow{S_1(u_{2\,k} - u_{1\,k+1})}
\left(\begin{smallmatrix}
u_{2\,k} & u_{1\,k} \medskip\\
u_{1\,k+1} & u_{2\,k+1}
\end{smallmatrix}
\right)\xleftarrow{S(u_{2\,k} - u_{2\,k+1})}
\left(\begin{smallmatrix}
u_{2\,k+1} & u_{1\,k} \medskip\\
u_{1\,k+1} & u_{2\,k}
\end{smallmatrix}
\right)\xleftarrow{S_1(u_{1\,k+1} - u_{2\,k+1})}
\left(\begin{smallmatrix}
u_{1\,k+1} & u_{1\,k} \medskip\\
u_{2\,k+1} & u_{2\,k}
\end{smallmatrix}
\right)
\end{align}
results in a needed permutations of parameters so that we obtain
\begin{align}\label{R}
R_{k+1 k}\left(\begin{smallmatrix}
u_{1\,k+1} & u_{1\,k} \medskip\\
u_{2\,k} & u_{2\,k+1}
\end{smallmatrix}
\right) = S_1(u_{1\,k+1} - u_{2\,k+1})\,S(u_{2\,k} - u_{2\,k+1})\,S_1(u_{2\,k} - u_{1\,k+1})
\end{align}
The product of R-operators
\begin{align}\nonumber
\Lambda_{v}\left(\begin{smallmatrix}
u_{1 N} & u_{1\,N-1} & \dots & u_{12}\medskip\\
u_{2\,N-1} & u_{2\,N-2} & \dots & u_{21}
\end{smallmatrix}
\right) = R_{N N-1}\left(\begin{smallmatrix}
u_{1\,N} & u_{1\,N-1} \medskip\\
u_{2\,N-1} & v
\end{smallmatrix}
\right)
R_{N-1 N-2}\left(\begin{smallmatrix}
u_{1\,N-1} & u_{1\,N-2} \medskip\\
u_{2\,N-2} & v
\end{smallmatrix}
\right)\cdots
R_{2 1}\left(\begin{smallmatrix}
u_{1 2} & u_{1 1} \medskip\\
u_{2 1} & v
\end{smallmatrix}
\right)
\end{align}
intertwines the monodromy matrices
\begin{align}\nonumber
T\left(\begin{smallmatrix}
u_{1 N} & u_{1\,N-1} & \dots
& u_{1 2} & u_{1 1} \medskip\\
v & u_{2\,N-1} & \dots
& u_{2 2} & u_{2 1}
\end{smallmatrix}
\right)\,\, \Lambda_{v}\left(\begin{smallmatrix}
u_{1 N} & u_{1\,N-1} & \dots & u_{12}\medskip\\
u_{2\,N-1} & u_{2\,N-2} & \dots & u_{21}
\end{smallmatrix}
\right) = \\
\Lambda_{v}\left(\begin{smallmatrix}
u_{1 N} & u_{1\,N-1} & \dots & u_{12}\medskip\\
u_{2\,N-1} & u_{2\,N-2} & \dots & u_{21}
\end{smallmatrix}
\right)
\,\, T\left(\begin{smallmatrix}
u_{1\, N} & u_{1\,N-1} & \dots & u_{1 2} & u_{11} \medskip\\
u_{2\, N-1} & u_{2\,N-2} & \dots & u_{2 1} & v
\end{smallmatrix}
\right)\,
\end{align}
and due to~(\ref{v}) it is equivalent to the following
intertwining relation for B-operators
\begin{align}\nonumber
B\left(\begin{smallmatrix}
u_{1 N} & u_{1\,N-1} & \dots & u_{1 2} & u_{1 1} \medskip\\
u_{2 N} & u_{2\,N-1} & \dots & u_{2 2} & u_{2 1}
\end{smallmatrix}
\right)\,\, \Lambda_{v}\left(\begin{smallmatrix}
u_{1 N} & u_{1\,N-1} & \dots & u_{12}\medskip\\
u_{2\,N-1} & u_{2\,N-2} & \dots & u_{21}
\end{smallmatrix}
\right) = \\
\Lambda_{v}\left(\begin{smallmatrix}
u_{1 N} & u_{1\,N-1} & \dots & u_{12}\medskip\\
u_{2\,N-1} & u_{2\,N-2} & \dots & u_{21}
\end{smallmatrix}
\right)
\,\, B\left(\begin{smallmatrix}
u_{1\, N} & u_{1\,N-1} & \dots & u_{1 2} & u_{11} \medskip\\
u_{2\, N-1} & u_{2\,N-2} & \dots & u_{2 1} & v
\end{smallmatrix}
\right)\,
\end{align}
The needed operator $W(U, V)$ is constructed step by step
\begin{align}\nonumber
W(U, V) = \Lambda_{v_{1}}\left(\begin{smallmatrix}
u_{1 N} & \dots & u_{12}\medskip\\
u_{2\,N-1} & \dots & u_{21}
\end{smallmatrix}
\right)\Lambda_{v_{2}}\left(\begin{smallmatrix}
u_{1 N} & \dots & u_{13}\medskip\\
u_{2\,N-2} & \dots & u_{21}
\end{smallmatrix}
\right) \cdots \Lambda_{v_{N-2}}\left(\begin{smallmatrix}
u_{1 N} & u_{1\, N-1} \medskip\\
u_{2 2} & u_{21}
\end{smallmatrix}
\right)
R_{N N-1}\left(\begin{smallmatrix}
u_{1\,N} & u_{1\,N-1} \medskip\\
u_{2 1} & v_{N-1}
\end{smallmatrix}
\right)
\end{align}
and intertwines the B-operators
\begin{align}\nonumber
B\left(\begin{smallmatrix}
u_{1 N} & u_{1\,N-1} & \dots & u_{1 2} & u_{1 1} \medskip\\
u_{2\, N} & u_{2\,N-1} & \dots & u_{2 2} & u_{2 1}
\end{smallmatrix}
\right)\, W(U, V) =
W(U, V)\, B\left(\begin{smallmatrix}
u_{1\, N} & u_{1\,N-1} & \dots & u_{1 2} & u_{1 1} \medskip\\
u_{2 1} & v_{N-1} & \dots & v_{2} & v_{1}
\end{smallmatrix}
\right)\,.
\end{align}
If $\Psi_0(x)$ is eigenfunction of $B(V)$
\begin{eqnarray}\label{want}
B(V)\Psi_0(x)  = -ip\,(u-q_1)\cdots (u-q_{N-1})\Psi_0(x)\,,
\end{eqnarray}
then $W(U, V)\Psi_0$ is eigenfunction of $B(U)$ with the same eigenvalues. Since $V$ is arbitrary, it is sufficient to construct only one eigenfunction of $B(V)$ and it will give rise to the family of eigenfunctions of $B(U).$
Let us consider such "particular" eigenfunction
$\Psi_0 = e^{i p x_N}.$\footnote{We remind that this function $e^{i p x_N + i \bar{p} \bar{x}_N}$ also contains antiholomorphic part, but it is not shown to simplify formulas.} Due to~(\ref{Laxsl2}) we have
\begin{align}\nonumber
T\! \left(\begin{smallmatrix}
u_{1N} & \dots & u_{12} & u_{11}\medskip\\
u_{21} & \dots & v_{2} & v_{1}
\end{smallmatrix}
\right)\,e^{ip x_N} = e^{ip x_N}\,
\left( \begin{array}{cc}
u_{1N}+1 +ip x_N & -ip\\
ip x_N^2 + (u_{21}-u_{1N}-1)x_N & u_{21} -ip x_N
\end{array}
\right) \\
\left( \begin{array}{cc}
u_{1\,N-1}+1 & 0\\
(v_{N-1}-u_{1\,N-1}-1)x_{N-1} & v_{N-1}
\end{array}
\right)\cdots
\left( \begin{array}{cc}
u_{11}+1 & 0\\
(v_{1}-u_{11}-1)x_1 & v_{1}
\end{array}
\right)
\end{align}
and therefore
$$
B(V) \Psi_0 = -ip\, v_{1}v_{2}\cdots v_{N-1} \, \Psi_0\,.
$$
Hence, if we put $v_{k} = u - q_k$ we arrive at desired form (\ref{want}).
Finally, eigenfunction $\Psi_p(q|x)$ satisfying
$$
B(U)\Psi_p(q|x) = -ip\,(u-q_1)\cdots(u-q_{N-1})\,\Psi_p(q|x)
$$
has the form
\begin{equation}
\Psi_p(q|x) = W(U,V)\,e^{ip x_N},
\end{equation}
where $v_{k} = u - q_k$ and $W(U, V)$ is constructed in an explicit form.
This formula presents known result for $\Psi_p(q|x)$~\cite{DKM,DM} in a little bit different form. The idea of this construction can be implemented to the case of $SL(3, \mathbb{C})$ magnet.

\section{Eigenfunctions of the operator $B(u)$ for the $SL(3, \mathbb{C})$ magnet.}

The initial expression for the operator $B(u)$ can be rewritten
in a useful matrix form
\begin{align}\label{B1}
B(u) = \left( T^1_3(u),\ T^2_3(u) \right) \left( \begin{array}{c} T^{12}_{13}(u+1)\\ T^{12}_{23}(u+1) \end{array} \right) = \\
\left( T^1_3(u),\ T^2_3(u) \right) \left( \begin{array}{cc}
T^{1}_{1}(u) & T^{2}_{1}(u) \\
T^{1}_{2}(u) & T^{2}_{2}(u)
\end{array}\right) \left( \begin{array}{c} T^{2}_{3}(u+1) \\
- T^{1}_{3}(u+1)  \end{array} \right)\,,
\end{align}
where in the last line we use relation
\begin{equation}\label{TT}
\left( \begin{array}{c} T^{12}_{13}(u+1)\\ T^{12}_{23}(u+1) \end{array} \right) = \left( \begin{array}{cc}
T^{1}_{1}(u) & T^{2}_{1}(u) \\
T^{1}_{2}(u) & T^{2}_{2}(u)
\end{array}\right) \left( \begin{array}{c} T^{2}_{3}(u+1) \\
- T^{1}_{3}(u+1)  \end{array} \right)\,
\end{equation}
which is simply matrix form of the expression for the quantum minors
\begin{align}\nonumber
T^{12}_{13}(u+1) = T^{1}_{1}(u)\,T^{2}_{3}(u+1) - T^{2}_{1}(u)\,T^{1}_{3}(u+1)\,, \\
T^{12}_{23}(u+1) = T^{1}_{2}(u)\,T^{2}_{3}(u+1) - T^{2}_{2}(u)\,T^{1}_{3}(u+1)\,.
\end{align}

\subsection{Example: $N=1.$}
\label{example}

To start with let us consider the simplest example of the one site $N=1$ when the monodromy matrix coincides with the L-operator:
$T^{i}_{j}(u) = L^{i}_{j}(\mathbf{u})$.
Substitution of the explicit matrix elements for the
L-operator~(\ref{Laxsl3}) gives
\begin{align}\label{LL}
\left( \begin{array}{c} L^{12}_{13}(u+1)\\ L^{12}_{23}(u+1) \end{array} \right) =
\left(  \footnotesize{\begin{array}{c|c}
u_1 + 2 + x\partial_x + y\partial_y &
\begin{array}{c}y\partial_z +\\x (x\partial_x + y\partial_y - z\partial_z + u_1 - u_2 +1)\end{array} \\ \hline
-\partial_x & u_2 + 1 -x\partial_x + z\partial_z
\end{array}}\right) \left(\footnotesize{ \begin{array}{c} -\partial_z - x\partial_y \\
\partial_y  \end{array}} \right)\,
\end{align}
and expression for the operator $B(u)$ explicitly reads
\begin{align}\nonumber
B(u) = -\left(\footnotesize{\partial_y ,\ \partial_z + x\partial_y }\right)
\left(  \footnotesize{\begin{array}{c|c}
u_1 + 2 + x\partial_x + y\partial_y &
\begin{array}{c}y\partial_z +\\x (x\partial_x + y\partial_y - z\partial_z + u_1 - u_2 +1)\end{array} \\ \hline
-\partial_x & u_2 + 1 -x\partial_x + z\partial_z
\end{array}}\right) \left(\footnotesize{ \begin{array}{c} -\partial_z - x\partial_y \\
\partial_y  \end{array}} \right)\,
\end{align}
Let us search for the eigenfunction of $B(u)$ of the form
\begin{equation}\label{Pss}
\Psi = e^{ip_1(y - x z) + ip_2 z }\varphi(x),
\end{equation}
where $\varphi$ is yet undefined function, and $p_{1,2}$ are parameters.
The direct calculation gives
\begin{align}\label{B1psi}
B(u)\Psi = -e^{ip_1(y - x z) + ip_2 z }\,
\left(\footnotesize{p_1 ,\ p_2 }\right)
\left(  \footnotesize{\begin{array}{cc}
u_1 + 2 + x\partial_x  &
x (x\partial_x  + u_1 - u_2 +1) \\
-\partial_x & u_2 + 1 -x\partial_x
\end{array}}\right) \left(\footnotesize{ \begin{array}{c} p_2 \\
-p_1  \end{array}} \right)\,\varphi(x)\,
\end{align}
Note that the matrix in the middle coincides up to
transposition and shift of the spectral parameter
$u \to u+1$ with the L-operator~(\ref{Laxsl2}) for $SL(2,\mathbb{C})$
\begin{align}\nonumber
L(u) = \left( \begin{array}{cc}
x\partial_x +u_1+1& -\partial_x\\
x\left(x\partial_x +u_1-u_2+1\right) & u_2 - x\partial_x
\end{array}
\right)
\end{align}
Of course it is possible to diagonalize the operator
\begin{align}\label{Bnew}
\left(\footnotesize{p_1 ,\ p_2 }\right)
L^{t}(u+1)
\left(\footnotesize{ \begin{array}{c} p_2 \\
-p_1  \end{array}} \right) = \left(\footnotesize{p_2 ,\ -p_1 }\right)
L(u+1)
\left(\footnotesize{ \begin{array}{c} p_1 \\
p_2  \end{array}} \right)
\end{align}
directly but we apply some trick which will be used
later in a general case of N sites.

Note that the matrix $E^{\boldsymbol{\sigma}}=\sum_{ij} E_{ij}e_{ji}$
coincides up to additive constant with the quadratic Casimir operator in the tensor product of representation $T^{\boldsymbol{\sigma}}$ and fundamental representation so that it commutes with the operators $T^{\boldsymbol{\sigma}}(g)\otimes g$
\begin{align}\nonumber
\left(T^{\boldsymbol{\sigma}}(g)\otimes g\right)
E^{\boldsymbol{\sigma}} = E^{\boldsymbol{\sigma}} \left(T^{\boldsymbol{\sigma}}(g)\otimes g\right)  \longrightarrow T^{\boldsymbol{\sigma}}(g)\,
E^{\boldsymbol{\sigma}}\,
T^{\boldsymbol{\sigma}}(g)^{-1} = g^{-1}\,E^{\boldsymbol{\sigma}}\,g
\end{align}
In terms of $L(u)$ it reads
\begin{equation}\label{ceq1}
T^{\boldsymbol{\sigma}}(g) L(u)
T^{\boldsymbol{\sigma}}(g)^{-1} = g^{-1}\,L(u) \,g,
\end{equation}
i.e. the matrix similarity transformation of $L$-operator  $L(u)\to g^{-1}\,L(u) \,g$ can be performed using the operator $T^{\boldsymbol{\sigma}}\left(g\right)$ acting on the quantum space.
The operator~(\ref{Bnew}) coincides with the
matrix element $\tilde{B}(u)$ of unitary transformed
monodromy matrix for one site:
\begin{equation}
\left(
\begin{array}{cc}
\tilde{A}(u) & \tilde{B}(u)\\ \tilde{C}(u) & \tilde{D}(u)
\end{array}\right) =
\left(
\begin{array}{cc}
p_2 & -p_1\\ 0 & p_2^{-1}
\end{array}\right)
\left(
\begin{array}{cc}
A(u) & B(u)\\ C(u) & D(u)
\end{array}\right)
\left(
\begin{array}{cc}
p_2^{-1} & p_1\\ 0 & p_2
\end{array}\right)\label{unit}
\end{equation}
Eq. (\ref{ceq1}) states that this transformation can be written as operator (which we will denote $\Omega$) acting in quantum space, i.e.
\begin{equation}
\tilde{B}(u)= \Omega\, B(u) \, \Omega^{-1}\,,
\end{equation}
and $\Omega$ is defined by (\ref{Tsl2}) with
$g = \left(\footnotesize{
\begin{array}{cc}
p_2^{-1} & p_1\\ 0 & p_2
\end{array}}\right)$:
\begin{align}
\Omega\phi(x)=[p_2 - p_1 x]^{u_2 - u_1 - 1}\,
\phi\left(\textstyle\frac{p_2^{-1} x}{p_2 - p_1 x_1}\right).
\end{align}
If $\varphi$ is eigenfunction of $B(u)$ then $\Omega \varphi$ is eigenfunction of $\tilde{B}(u).$ But eigenfunctions of $SL(2,\mathbb{C})$ operators  $B(u) = -\partial_x$ and $\bar{B}(u) = -\bar{\partial}_x$ are simply exponents\footnote{We understand all exponent here as the product of holomorphic and antiholomorphic part, i.e. $e^{ipx}$ stands for $e^{ipx+i\bar{p}\bar{x}}.$}
$$
\varphi_p(x) = e^{ipx}.
$$
This system of functions is orthogonal and complete
\begin{eqnarray}\label{exp}
\int d^2 x\, \overline{\varphi_p(x)} \varphi_{p'}(x) =  \pi^2 \delta^{2}(\vec{p}-\vec{p}^{\,\prime})\ \ ;\ \
\int \frac{d^2 p}{\pi^2}\,\, \overline{\varphi_p(x)}\varphi_p(x') =  \delta^{2}(\vec{x}-\vec{x}^{\,\prime}).
\end{eqnarray}
The function $\Psi_{p_1 p_2 p}(x, y, z)$
\begin{equation}
\Psi_{p_1 p_2 p}(x, y, z) = e^{ip_1(y-xz) +i p_2 z} \Omega\, e^{ipx}
= [p_2 - p_1 x]^{u_2 - u_1 - 1}\,
e^{ip_1(y-xz) +ip_2 z+\frac{ip}{p_2}\frac{x}{p_2 - p_1x}}
\end{equation}
satisfy
\begin{eqnarray}
B(u)\Psi_{p_1 p_2 p}(x, y, z) =  ip\,\Psi_{p_1 p_2 p}(x, y, z)\,,\\
E_{31}\Psi_{p_1 p_2 p}(x, y, z) =
(-\partial_y)\Psi_{p_1 p_2 p}(x, y, z) = -ip_1 \Psi_{p_1 p_2 p}(x, y, z)\,,\\
E_{32}\Psi_{p_1 p_2 p}(x, y, z)  =  (-\partial_z- x\partial_y)\Psi_{p_1 p_2 p}(x, y, z) = -ip_2 \Psi_{p_1 p_2 p}(x, y, z)\,.
\end{eqnarray}
The corresponding orthogonality and completeness relations
\begin{equation}
\int d^2x\, d^2y\, d^2z\, \overline{\Psi_{p_1p_2p}(x, y, z)} \Psi_{p'_1 p'_2 p'}(x, y, z) =
\pi^6\, \delta^{2}(\vec{p}_1 - \vec{p}_1^{\,\prime})\,\delta^{2}(\vec{p}_2 - \vec{p}_2^{\,\prime})\,\delta^{2}(\vec{p}-\vec{p}^{\,\prime})
\end{equation}
\begin{equation}
\int \frac{d^2p_1}{\pi^2}\,\frac{d^2p_2}{\pi^2}\,
\frac{d^2p}{\pi^2}\, \overline{\Psi_{p_1p_2p}(x, y, z)}
\Psi_{p_1p_2p}(x', y', z') =
\delta^{2}(\vec{x} - \vec{x}^{\,\prime})\,\delta^{2}(\vec{y} - \vec{y}^{\,\prime})\,\delta^{2}(\vec{z}-\vec{z}^{\,\prime})
\end{equation}
can be proven with the help of (\ref{exp}).
Now we are going to the general situation of N sites.

\subsection{Permutation of parameters}

For the group $SL(3, \mathbb{C})$ the $L-$operator
depends on three parameters
\begin{align}
L_k(\mathbf{u}_k) = L_k\left(\begin{smallmatrix}
u_{1 k}\smallskip\\
u_{2 k}\smallskip\\
u_{3 k}
\end{smallmatrix}
\right)\ \ ;\ \
u_{i k} = u - \sigma_{i}^{(k)} + \delta_k\ ;\ i=1\,,2\,,3\,,
\end{align}
and, similar to $SL(2, \mathbb{C})$ case we introduce matrix of parameters
\begin{equation}
U = \left( \begin{matrix}
u_{1n} & u_{1\,n-1} & \dots & u_{1\,k+1} & u_{1 k}\\
u_{2n} & u_{2\,n-1} & \dots & u_{1\,k+1} & u_{2 k}\\
u_{3n} & u_{3\,n-1} & \dots & u_{1\,k+1} & u_{3 k}
\end{matrix}
\right)
\end{equation}
for the monodromy matrix from site k to site n
\begin{equation}
T(U) = L_n(\mathbf{u}_n)L_{n-1}(\mathbf{u}_{n-1})
\cdots L_{k+1}(\mathbf{u}_{k+1}) L_k(\mathbf{u}_k)\,.
\end{equation}
An essential role will be played by operators $S_{1}(\mathbf{u}_k)$ and $S_{2}(\mathbf{u}_k)$~(\ref{S12}) which perform the parameters permutations
$u_{1 k}\rightleftarrows u_{2 k}$ and $u_{2 k}\rightleftarrows u_{3 k}$
inside the L-operator at k-th site
\begin{equation}
L_k \left(\begin{smallmatrix}
u_{1 k}\smallskip\\
u_{2 k}\smallskip\\
u_{3 k}
\end{smallmatrix}
\right)\,
S_1(\mathbf{u}_k) = S_1(\mathbf{u}_k)\,
L_k
\left(\begin{smallmatrix}
u_{2 k}\smallskip\\
u_{1 k}\smallskip\\
u_{3 k}
\end{smallmatrix}
\right) \ \ ;\
L_k \left(\begin{smallmatrix}
u_{1 k}\smallskip\\
u_{2 k}\smallskip\\
u_{3 k}
\end{smallmatrix}
\right)\,
S_2(\mathbf{u}_k) = S_2(\mathbf{u}_k)\,
L_k
\left(\begin{smallmatrix}
u_{1 k}\smallskip\\
u_{3 k}\smallskip\\
u_{2 k}
\end{smallmatrix}
\right) \,,
\label{s2l}
\end{equation}
and operator $S(\mathbf{u}_{k+1}\,,\mathbf{u}_k)$ which interchanges parameters
$u_{1\,k+1}\rightleftarrows u_{3\,k}$ inside the product of L-operators
at two adjacent sites
\begin{gather}
T\left(\begin{smallmatrix}
u_{1\,k+1} & u_{1\,k} \smallskip\\
u_{2\,k+1} & u_{2\,k} \smallskip\\
u_{3\,k+1} & u_{3\,k}
\end{smallmatrix}
\right) = L_{k+1}(\mathbf{u}_{k+1}) L_k(\mathbf{u}_k)\,, \medskip \\
T\left(\begin{smallmatrix}
u_{1\,k+1} & u_{1\,k} \smallskip\\
u_{2\,k+1} & u_{2\,k} \smallskip\\
u_{3\,k+1} & u_{3\,k}
\end{smallmatrix}
\right)\,
S(\mathbf{u}_{k+1}\,,\mathbf{u}_k) = S(\mathbf{u}_{k+1}\,,\mathbf{u}_k)\,T\left(\begin{smallmatrix}
u_{3\,k} & u_{1\,k} \smallskip\\
u_{2\,k+1} & u_{2\,k} \smallskip\\
u_{3\,k+1} & u_{1\,k+1}
\end{smallmatrix}
\right)\,.
\label{s3l}
\end{gather}
Direct calculation shows that it is again multiplication operator~\cite{DM3}
\begin{equation}
S(\mathbf{u}_{k+1}\,,\mathbf{u}_k) = S(u_{3 k}-u_{1\, k+1}) =
\left[ y_{k+1} - y_{k} - z_{k}(x_{k+1}-x_{k}) \right]^{u_{3 k}-u_{1\, k+1}}
\end{equation}
Clearly, $[S_1(\mathbf{u}_k)\,,L_i(\mathbf{u}_i)] = 0 $ for $i \ne k$
and $[S(\mathbf{u}_{k+1}\,,\mathbf{u}_k)\,,L_{i}(\mathbf{u}_{i})] = 0$ for $i \ne k, k+1$ so that its commutation relations with the complete monodromy matrix
from the first site to the N-th site simply mimics considered local commutation relations.
These three operators can be used to perform any permutation of elements in $U.$

\subsection{Parameter dependence of $B(U).$}

From the explicit form of $SL(3, \mathbb{C})$ $L-$operator (\ref{Laxsl3}) we see that $L_j^i(u)$ does not depend on $u_{1,2,3}$ for $i< j$ and $L^i_j(u) = L^i_j(u_j, \dots , u_i)$ for $i \ge j.$ For quantum minors of $L-$operator then $L_{j_1 j_2}^{i_1 i_2}(u) = L_{j_1 j_2}^{i_1 i_2}(u_{j_1},\dots u_{i_2})$ if $i_1 < i_2$ and $j_1 < j_2$ (it can be always fulfilled since is $L^{i_1 i_2}_{j_1 j_2}(u)$ antisymmetric under the permutation of $i_1,\ i_2$ and $j_1,\ j_2,$ see sec. 2).

The quantum minor of monodromy matrix can be expressed in terms of quantum minors of corresponding $L-$operators~\cite{MNO,M}:
\begin{equation}\label{coprod}
T^{i_1i_2}_{j_1 j_2}(u) = \sum\limits_{a_1 < a_2 , b_1<b_2 , \ldots}
L_N(u)^{i_1\ i_2}_{a_1 a_2}\, L_{N-1}(u)^{a_1 a_2}_{b_1 b_2}
\cdots L_1(u)^{c_1 c_2}_{j_1 j_2}
\end{equation}
which is similar to definition (\ref{Top}) written in terms of matrix elements:
\begin{equation}\label{coprod1}
T^{i}_{j}(u) = \sum\limits_{a,b\ldots} L_N(u)^{i}_{a}\, L_{N-1}(u)^{a}_{b}\dots L_1(u)^{c}_{j}.
\end{equation}
From (\ref{coprod}) we conclude that $T^{i_1 i_2}_{j_1 j_2}(u)$ does not depend on parameters $u_{j1}$ for $j< j_1$ and $u_{iN}$ for $i > i_2.$
Hence $T^{12}_{23}(u)$ and $T^{12}_{13}(u)$ does not depend on parameter $u_{3N}.$ The same is valid for $T^2_3(u)$ and $T^1_3(u)$ and we conclude that the whole
operator $B(u)$~(\ref{B1}) does not depend on $u_{3N}.$ This is in the full analogy with $SL(2, \mathbb{C})$ case.

Using similar idea as in (\ref{idea}) we can construct operator $W(U, V)$ which satisfy
\begin{equation}
B(U)\, W(U, V) = W(U, V)\, B(V)\,,\label{master}
\end{equation}
i.e. it changes the parameter matrix $U$ to the parameter matrix $V$
\begin{equation}
U = \left( \begin{matrix}
u_{1N} & \dots & u_{1 2} & u_{1 1}\\
u_{2N} & \dots & u_{2 2} & u_{2 1}\\
u_{3N} & \dots & u_{3 2} & u_{3 1}
\end{matrix}
\right)\ \ \longrightarrow\ \ \
V = \left( \begin{matrix}
v_{1N} & \dots & v_{1 2} & v_{1 1}\\
v_{2N} & \dots & v_{2 2} & v_{2 1}\\
u_{1N} & \dots & u_{1 2} & u_{1 1}
\end{matrix}
\right)
\end{equation}
containing arbitrary parameters in the first two rows. We shall suppose that $v_{ik}$ are linear functions of the spectral parameter $u$. The reason is that all operators $S_1\,,S_2$ and $S$ should not depend on $u$ to satisfy (\ref{s2l}), (\ref{s3l}) with $L(u)$ and $L(u+1).$

The operator $W(U, V)$ can be constructed from the elementary intertwining operators in a many equivalent ways.
We present the construction which is the direct generalization of the ones used for the case $SL(2,\mathbb{C})$ in the section~\ref{Bsl2} and the whole transformation will be performed in a two steps
$$
U = \left( \begin{smallmatrix}
u_{1N} & \dots & u_{1 2} & u_{1 1}\smallskip\\
u_{2N} & \dots & u_{2 2} & u_{2 1}\smallskip\\
u_{3N} & \dots & u_{3 2} & u_{3 1}
\end{smallmatrix}
\right)\ \longrightarrow\ \
V_1 = \left( \begin{smallmatrix}
v_{1N} & \dots & v_{1 2} & v_{1 1}\smallskip\\
u_{1N} & \dots & u_{1 2} & u_{1 1}\smallskip\\
u_{2N} & \dots & u_{2 2} & u_{2 1}
\end{smallmatrix}
\right)\ \longrightarrow\ \
V = \left( \begin{smallmatrix}
v_{1N} & \dots & v_{1 2} & v_{1 1}\smallskip\\
v_{2N} & \dots & v_{2 2} & v_{2 1}\smallskip\\
u_{1N} & \dots & u_{1 2} & u_{1 1}
\end{smallmatrix}
\right)
$$
The main building blocks are operators $R_{k+1 k}$
each of them interchanges parameters
$u_{3\,k+1}\rightleftarrows u_{3\,k}$ at two adjacent sites
\begin{gather}\nonumber
T\left(\begin{smallmatrix}
\dots & u_{1\,k+1} & u_{1\,k} & \dots \smallskip\\
\dots & u_{2\,k+1} & u_{2\,k} & \dots \smallskip\\
\dots & u_{3\,k+1} & u_{3\,k} & \dots
\end{smallmatrix}
\right) R_{k+1 k}\left(\begin{smallmatrix}
u_{1\,k+1} & u_{1\,k} \smallskip\\
u_{2\,k+1} & u_{2\,k} \smallskip \\
u_{3\,k} & u_{3\,k+1}
\end{smallmatrix}
\right) = R_{k+1 k}\left(\begin{smallmatrix}
u_{1\,k+1} & u_{1\,k} \smallskip\\
u_{2\,k+1} & u_{2\,k} \smallskip \\
u_{3\,k} & u_{3\,k+1}
\end{smallmatrix}
\right) T\left(\begin{smallmatrix}
\dots & u_{1\,k+1} & u_{1\,k} & \dots \smallskip\\
\dots & u_{2\,k+1} & u_{2\,k} & \dots \smallskip\\
\dots & u_{3\,k} & u_{3\,k+1} & \dots
\end{smallmatrix}
\right)
\end{gather}
Note that the parameters in R-matrix mimic exactly parameters in
the monodromy matrix in the right hand side of the considered relation.
The chain of the elementary transpositions
\begin{align}\nonumber
\left(\begin{smallmatrix}
u_{1\,k+1} & u_{1\,k} \smallskip\\
u_{2\,k+1} & u_{2\,k}\smallskip\\
u_{3\,k} & u_{3\,k+1}
\end{smallmatrix}
\right)\xleftarrow{S_{2}(u_{3\,k} - u_{2\,k+1})}
\left(\begin{smallmatrix}
u_{1\,k+1} & u_{1\,k} \smallskip\\
u_{3\,k} & u_{2\,k}\smallskip\\
u_{2\,k+1} & u_{3\,k+1}
\end{smallmatrix}
\right)
\xleftarrow{S_1(u_{3\,k} - u_{1\,k+1})}
\left(\begin{smallmatrix}
u_{3\,k} & u_{1\,k} \smallskip\\
u_{1\,k+1} & u_{2\,k}\smallskip\\
u_{2\,k+1} & u_{3\,k+1}
\end{smallmatrix}
\right)\xleftarrow{S(u_{3 k} -u_{3\,k+1})}\\
\nonumber
\left(\begin{smallmatrix}
u_{3\,k+1} & u_{1\,k} \smallskip\\
u_{1\,k+1} & u_{2\,k}\smallskip\\
u_{2\,k+1} & u_{3\,k}
\end{smallmatrix}
\right)\xleftarrow{S_1(u_{1\,k+1} - u_{3\,k+1})}
\left(\begin{smallmatrix}
u_{1\,k+1} & u_{1\,k} \smallskip\\
u_{3\,k+1} & u_{2\,k}\smallskip\\
u_{2\,k+1} & u_{3\,k}
\end{smallmatrix}
\right)
\xleftarrow{S_2(u_{2\,k+1} - u_{3\,k+1})}
\left(\begin{smallmatrix}
u_{1\,k+1} & u_{1\,k} \smallskip\\
u_{2\,k+1} & u_{2\,k}\smallskip\\
u_{3\,k+1} & u_{3\,k}
\end{smallmatrix}
\right)
\end{align}
results in a needed permutations of parameters so that we have
\begin{align}\nonumber
R_{k+1 k}\left(\begin{smallmatrix}
u_{1\,k+1} & u_{1\,k} \smallskip\\
u_{2\,k+1} & u_{2\,k} \smallskip \\
u_{3\,k} & u_{3\,k+1}
\end{smallmatrix}
\right) = S_2(u_{2\,k+1} - u_{3\,k+1})\,
S_1(u_{1\,k+1} - u_{3\,k+1})\\ \label{R3}
S(u_{3\,k} - u_{3\,k+1})\,S_1(u_{3\,k} - u_{1\,k+1})
\,S_2(u_{3\,k} - u_{2\,k+1})
\end{align}
The product of R-operators
\begin{align}\nonumber
\Lambda_{v}\left(\begin{smallmatrix}
u_{1 N} & u_{1\,N-1} & \dots & u_{12}\smallskip\\
u_{2 N} & u_{2\,N-1} & \dots & u_{22}\smallskip\\
u_{3\,N-1} & u_{3\,N-2} & \dots & u_{31}
\end{smallmatrix}
\right) = R_{N N-1}\left(\begin{smallmatrix}
u_{1\,N} & u_{1\,N-1} \smallskip\\
u_{2\,N} & u_{2\,N-1} \smallskip\\
u_{3\,N-1} & v
\end{smallmatrix}
\right)
R_{N-1 N-2}\left(\begin{smallmatrix}
u_{1\,N-1} & u_{1\,N-2} \smallskip\\
u_{2\,N-1} & u_{2\,N-2} \smallskip\\
u_{3\,N-2} & v
\end{smallmatrix}
\right)\cdots
R_{2 1}\left(\begin{smallmatrix}
u_{1 2} & u_{1 1} \smallskip\\
u_{2 2} & u_{2 1} \smallskip\\
u_{3 1} & v
\end{smallmatrix}
\right)
\end{align}
intertwines the following monodromy matrices
\begin{align}\nonumber
\nonumber
T\left(\begin{smallmatrix}
u_{1 N} & u_{1\,N-1} & \dots & u_{1 2} & u_{1 1} \smallskip\\
u_{2 N} & u_{2\,N-1} & \dots & u_{2 2} & u_{2 1} \smallskip\\
v & u_{3\,N-1} & \dots & u_{3 2} & u_{3 1}
\end{smallmatrix}
\right)\,\, \Lambda_{v}\left(\begin{smallmatrix}
u_{1 N} & u_{1\,N-1} & \dots & u_{12}\smallskip\\
u_{2 N} & u_{2\,N-1} & \dots & u_{22}\smallskip\\
u_{3\,N-1} & u_{3\,N-2} & \dots & u_{31}
\end{smallmatrix}
\right) = \\
\Lambda_{v}\left(\begin{smallmatrix}
u_{1 N} & u_{1\,N-1} & \dots & u_{12}\smallskip\\
u_{2 N} & u_{2\,N-1} & \dots & u_{22}\smallskip\\
u_{3\,N-1} & u_{3\,N-2} & \dots & u_{31}
\end{smallmatrix}
\right)\,T\left(\begin{smallmatrix}
u_{1\, N} & u_{1\,N-1} & \dots & u_{1 2} & u_{11} \smallskip\\
u_{2\, N} & u_{2\,N-1} & \dots & u_{2 2} & u_{21} \smallskip\\
u_{3\, N-1} & u_{3\,N-2} & \dots & u_{3 1} & v
\end{smallmatrix}
\right)\,\,
\end{align}
and then we apply the appropriate intertwining operators
$$
S_{21}\left(\begin{smallmatrix}
v\smallskip\\
u_{11}\smallskip\\
u_{21}
\end{smallmatrix}
\right) \equiv S_2(u_{21}-v)S_1(u_{11}-v)
$$ at the first site
\begin{align}\nonumber
T\left(\begin{smallmatrix}
u_{1\, N} & u_{1\,N-1} & \dots & u_{1 2} & u_{11} \smallskip\\
u_{2\, N} & u_{2\,N-1} & \dots & u_{2 2} & u_{21} \smallskip\\
u_{3\, N-1} & u_{3\,N-2} & \dots & u_{3 1} & v
\end{smallmatrix}
\right)\,S_{21}\left(\begin{smallmatrix}
v\smallskip\\
u_{11}\smallskip\\
u_{21}
\end{smallmatrix}
\right) = S_{21}\left(\begin{smallmatrix}
v\smallskip\\
u_{11}\smallskip\\
u_{21}
\end{smallmatrix}
\right)\,T\left(\begin{smallmatrix}
u_{1\, N} & u_{1\,N-1} & \dots & u_{1 2} & v \smallskip\\
u_{2\, N} & u_{2\,N-1} & \dots & u_{2 2} & u_{11} \smallskip\\
u_{3\, N-1} & u_{3\,N-2} & \dots & u_{3 1} & u_{21}
\end{smallmatrix}
\right)\,
\end{align}
where again the parameters in operator $S_{21}$ mimic exactly parameters in
the last column of the monodromy matrix in the right hand side of the considered relation.

The operator $W(U, V_1)$ is constructed step by step
\begin{align}\nonumber
W(U, V_1) = \Lambda_{v_{11}}\left(\begin{smallmatrix}
u_{1 N} & \dots & u_{12}\smallskip\\
u_{2 N} & \dots & u_{22}\smallskip\\
u_{3\,N-1} & \dots & u_{31}
\end{smallmatrix}
\right)\,
S_{21}\left(\begin{smallmatrix}
v_{11}\smallskip\\
u_{11}\smallskip\\
u_{21}
\end{smallmatrix}
\right)\,
\Lambda_{v_{12}}\left(\begin{smallmatrix}
u_{1 N} & \dots & u_{13}\smallskip\\
u_{2 N} & \dots & u_{23}\smallskip\\
u_{3\,N-2} & \dots & u_{31}
\end{smallmatrix}
\right)\,
S_{21}\left(\begin{smallmatrix}
v_{12}\smallskip\\
u_{12}\smallskip\\
u_{22}
\end{smallmatrix}
\right)
\cdots \\
\nonumber
\Lambda_{v_{1\,N-2}}\left(\begin{smallmatrix}
u_{1 N} & u_{1\, N-1} \smallskip\\
u_{2 N} & u_{2\, N-1} \smallskip\\
u_{3 2} & u_{31}
\end{smallmatrix}
\right)\,
S_{21}\left(\begin{smallmatrix}
v_{1\,N-2}\smallskip\\
u_{1\,N-2}\smallskip\\
u_{2\,N-2}
\end{smallmatrix}
\right)
R_{N N-1}\left(\begin{smallmatrix}
u_{1\,N} & u_{1\,N-1} \smallskip\\
u_{2\,N} & u_{2\,N-1} \smallskip\\
u_{3 1} & v_{1\,N-1}
\end{smallmatrix}
\right)\,
S_{21}\left(\begin{smallmatrix}
v_{1\,N-1}\smallskip\\
u_{1\,N-1}\smallskip\\
u_{2\,N-1}
\end{smallmatrix}
\right)
S_{21}\left(\begin{smallmatrix}
v_{1\,N}\smallskip\\
u_{1\,N}\smallskip\\
u_{2\,N}
\end{smallmatrix}
\right)
\end{align}
and intertwines the operators $B(V_1)$ and $B(U)$
\begin{align}\nonumber
B\left(\begin{smallmatrix}
u_{1\, N} & u_{1\,N-1} & \dots & u_{1 2} & u_{1 1} \smallskip\\
u_{2\, N} & u_{2\,N-1} & \dots & u_{2 2} & u_{2 1} \smallskip\\
u_{3\, N} & u_{3\,N-1} & \dots & u_{3 2} & u_{3 1}
\end{smallmatrix}
\right) \, W(U, V_1) =
W(U, V_1)\, B\left(\begin{smallmatrix}
v_{1\, N} & v_{1\,N-1} & \dots & v_{1 2} & v_{1 1} \smallskip\\
u_{1\, N} & u_{1\,N-1} & \dots & u_{1 2} & u_{1 1} \smallskip\\
u_{2\, N} & u_{2\,N-1} & \dots & u_{2 2} & u_{2 1}
\end{smallmatrix}
\right)\,.
\end{align}
In a similar way one constructs the operator $W(V_1, V)$
\begin{align}\nonumber
W(V_1, V) = \Lambda_{v_{21}}\left(\begin{smallmatrix}
v_{1 N} & \dots & v_{12}\smallskip\\
u_{1 N} & \dots & u_{12}\smallskip\\
u_{2\,N-1} & \dots & u_{21}
\end{smallmatrix}
\right)\,S_{2}(u_{21} - v_{21})\,
\Lambda_{v_{22}}\left(\begin{smallmatrix}
v_{1 N} & \dots & v_{13}\smallskip\\
u_{1 N} & \dots & u_{13}\smallskip\\
u_{2\,N-2} & \dots & u_{21}
\end{smallmatrix}
\right)\,S_{2}(u_{22} - v_{22})
\cdots \\
\nonumber
\Lambda_{v_{2\,N-2}}\left(\begin{smallmatrix}
v_{1 N} & v_{1\, N-1} \smallskip\\
u_{1 N} & u_{1\, N-1} \smallskip\\
u_{2 2} & u_{21}
\end{smallmatrix}
\right)\,S_{2}(u_{2\,N-2} - v_{2\,N-2})\,
R_{N N-1}\left(\begin{smallmatrix}
u_{1\,N} & u_{1\,N-1} \smallskip\\
u_{2\,N} & u_{2\,N-1} \smallskip\\
u_{3 1} & v_{1\,N-1}
\end{smallmatrix}
\right)
\,S_{2}(u_{2\,N-1} - v_{2\,N-1})
\,S_{2}(u_{2\,N} - v_{2\,N})
\end{align}
which intertwines the operators $B(V)$ and $B(V_1)$
\begin{align}\nonumber
B\left(\begin{smallmatrix}
v_{1\, N} & v_{1\,N-1} & \dots & v_{1 2} & v_{1 1} \smallskip\\
u_{1\, N} & u_{1\,N-1} & \dots & u_{1 2} & u_{1 1} \smallskip\\
u_{2\, N} & u_{2\,N-1} & \dots & u_{2 2} & u_{2 1}
\end{smallmatrix}
\right) \, W(V_1, V) = W(V_1, V)\, B\left(\begin{smallmatrix}
v_{1\, N} & v_{1\,N-1} & \dots & v_{1 2} & v_{1 1} \smallskip\\
v_{2\, N} & v_{2\,N-1} & \dots & v_{2 2} & v_{2 1} \smallskip\\
u_{1\, N} & u_{1\,N-1} & \dots & u_{1 2} & u_{1 1}
\end{smallmatrix}
\right)\,.
\end{align}
The needed operator $W(U, V)$ is the product $W(U, V) = W(U, V_1) W(V_1, V)$

Again, it is sufficient to find one eigenfunction $\Psi$ of $B(V)$ which eigenvalues depend on $v_{ik},$ and it will give rise to the set $W(U, V)\Psi$ of eigenfunctions of $B(U).$ This particular eigenfunction $\Psi$ can be found in recursive way by the reduction to $SL(2, \mathbb{C})$ case. This construction will be described in the next section.

\subsection{Reduction to the $SL(2, \mathbb{C})$ case}

Let us search the eigenfunction of $B(V)$
\begin{equation}\label{B2}
B(V) = \left( T^1_3(V),\ T^2_3(V) \right) \left( \begin{array}{c} T^{12}_{13}(V+1)\\ T^{12}_{23}(V+1) \end{array} \right).
\end{equation}
in the form
\begin{equation}\label{Pss}
\Psi = e^{ip_1(y_1 - x_1 z_1) + ip_2 z_1 }\varphi(x_1, x_2, \dots , x_N).
\end{equation}
Note that in the matrix $V$ all matrix elements $v_{ik}$ are linear function of the spectral parameter $u$ and we shall use compact notation $V+1$
for the matrix with matrix elements $v_{ik}+1$ which corresponds to the shift of the spectral parameter $u \to u+1$.

{\bf Proposition 1.} {\it The action of the column in the rhs of (\ref{B2}) on $\Psi$ is given by the formula}
\begin{align}\label{act1}
\left( \begin{array}{c} T^{12}_{13}(V+1)\\ T^{12}_{23}(V+1) \end{array} \right)\, \Psi = \prod\limits_{k=2}^N (v_{1k}+2)(v_{2k}+2)\, e^{ip_1(y_1 - x_1 z_1) +i p_2 z_1} \\ \nonumber
\left(  \footnotesize{\begin{array}{cc}
v_{11} + 2 + x_1\partial_{x_1}  &
x_1 (x_1\partial_{x_1}  + v_{11} - v_{21} +1) \\
-\partial_{x_1} & v_{21} + 1 -x_1\partial_{x_1}
\end{array}}\right) \left( \begin{array}{c} -ip_2 \\ ip_1  \end{array} \right)\, \varphi
\end{align}

{\bf Proof.} For the minors of $L-$operators for the sites $k=2, \dots, N$ one has (see~(\ref{LL}))
$$
L_k(\mathbf{v}_k+1)^{12}_{23}\,\Psi = L_k(\mathbf{v}_k+1)^{12}_{13}\,\Psi = 0;
\quad L_k(\mathbf{v}_k+1)^{12}_{12}\, \Psi = (v_{1k}+2)(v_{2k}+2)\Psi \,,
$$
and at the first site we obtain
\begin{align}\nonumber
\left( \begin{array}{c}
L_{1}(\mathbf{v}_1+1)^{12}_{13}\\
L_{1}(\mathbf{v}_1+1)^{12}_{23} \end{array} \right)\,\Psi =
e^{ip_1(y_1 - x_1 z_1) + ip_2 z_1 }
\left(  \footnotesize{\begin{array}{cc}
v_{11} + 2 + x_1\partial_{x_1}  &
x_1 (x_1\partial_{x_1}  + v_{11}-v_{21}+1) \\
-\partial_{x_1} & v_{21} + 1 -x_1\partial_{x_1}
\end{array}}\right)\left(\footnotesize{ \begin{array}{c} -ip_2 \\
ip_1  \end{array}} \right)\varphi\,
\end{align}
and this formula is very similar to~(\ref{B1psi}). Using (\ref{coprod}) one
arrives at desired formula (\ref{act1}). $\Box$

In each line of the r.h.s. of (\ref{act1}) we have again the function of the form (\ref{Pss}).

{\bf Proposition 2.} { \it The following formula holds:}

\begin{align}\label{act2}
\left( \begin{array}{c} T^1_3(V)\\ T^2_3(V)\end{array} \right)\Psi =
e^{ip_1(y_1 - x_1 z_1) +i p_2 z_1 }
\left(  \footnotesize{\begin{array}{cc}
v_{1N} + 2 + x_N\partial_{x_N}  &
-\partial_{x_N} \\
\nonumber
x_N (x_{N}\partial_{x_N}  + v_{1N}-v_{2N}+1) & v_{2N} + 1 -x_N\partial_{x_N}
\end{array}}\right)\, \\
\cdots \left(  \footnotesize{\begin{array}{cc}
v_{12} + 2 + x_2\partial_{x_2}  &
-\partial_{x_2} \\
x_2 (x_2\partial_{x_2}  + v_{12}-v_{22}+1) & v_{22} + 1 -x_2\partial_{x_2}
\end{array}}\right)\left(\footnotesize{ \begin{array}{c} -ip_1 \\ -ip_2  \end{array}} \right)
\varphi\,
\end{align}

{\bf Proof.} For the $L-$operators for the sites $k=2, \dots, N$ all matrix elements are acting on the function which depends on the $x-$variables only and we obtain
\begin{align}
\nonumber
L(\mathbf{u})\varphi(x) =
\left( \footnotesize{ \begin{array}{c|c|c} u_1 + 2 + x\partial_x  & -
\partial_x & 0\\
x (x\partial_x + u_1 - u_2 +1)
& u_2 + 1 -x\partial_x & 0\\ \hline
y(x\partial_x + u_1 - u_3 +2)-xz(u_2 - u_3 +1)&
- y\partial_x + z(u_2 - u_3 + 1)& u_3
\end{array}}\right)\varphi(x) \,.
\end{align}
Being substituted in (\ref{coprod1}) together with
$$
L_1(\mathbf{v}_1)^1_3\, \Psi = -ip_1\Psi;\quad
L_1(\mathbf{v}_1)^2_3\, \Psi = -ip_2 \Psi,
$$
it gives the desired expression (\ref{act2}).$\Box$\\
Combining results of two Propositions we get the formula for the action of $B(V)$ on $\Psi:$
\begin{equation}\label{Bact}
B(V)\Psi = -e^{ip_1(y_1 - x_1 z_1) +ip_2 z_1}\prod\limits_{k=2}^N (v_{1k}+2)(v_{2k}+2)\,\left[\left( p_1, \ p_2 \right)\left( \begin{array}{cc} a(V) & c(V) \\ b(V) & d(V) \end{array} \right) \left( \begin{array}{c} p_2 \\ -p_1 \end{array} \right)\right]\varphi,
\end{equation}
where matrix in the middle is the transposed monodromy matrix for the $SL(2,\mathbb{C})$ invariant spin chain with shifted spectral parameter and sites ordered as $ 2, \dots, N, 1$ from the right to the left
\begin{align}
\left( \begin{array}{cc} a(V) & c(V) \\ b(V) & d(V) \end{array} \right) = \left( \begin{array}{cc} a(V) & b(V) \\ c(V) & d(V) \end{array} \right)^t =
\left[L_1(\mathbf{v}_1+1)L_N(\mathbf{v}_N+1)\,\cdots\, L_2(\mathbf{v}_2+1)\right]^t\,,
\end{align}
where
\begin{align}\nonumber
L_k(\mathbf{v}_k+1) = \left( \begin{array}{cc}
x_k\partial_{x_k} +v_{1k}+2& -\partial_{x_k}\\
x_k\left(x_k\partial_{x_k} +v_{1k}-v_{2k}+1\right) &
v_{2k} +1 - x_k\partial_{x_k}
\end{array}
\right)
\end{align}
As a result, the function $\Psi$ (\ref{Pss}) is eigenfunction of $B(V)$ if and only if the function $\varphi,$ depending only on one variable $x_k$ in each site, is the eigenfunction of the operator
\begin{align}\nonumber
\left( p_1\,,p_2 \right)\left( \begin{array}{cc} a(V) & c(V) \\ b(V) & d(V) \end{array} \right)\left( \begin{array}{c} p_2 \\ -p_1 \end{array} \right) = \left( p_2\,,-p_1 \right)\left( \begin{array}{cc} a(V) & b(V) \\ c(V) & d(V) \end{array} \right) \left( \begin{array}{c} p_1 \\ p_2 \end{array} \right)
\end{align}

Such function $\varphi$ will be constructed with the use of
eigenfunctions of $b(V)$ in the next section.

\subsection{Unitary transformations of monodromy matrix}

We shall use the $SL(3,C)$ invariance of the L-operator
\begin{equation}\label{ceq}
T^{\boldsymbol{\sigma}}(g) L(u)
T^{\boldsymbol{\sigma}}(g)^{-1} = g^{-1}\,L(u) \,g
\end{equation}
in a closed analogy with section~\ref{example}.
Right-hand side of (\ref{Bact}) in square brackets contain matrix element $\tilde{b}(u)$ of unitary transformed monodromy matrix:
\begin{equation}
\left(
\begin{array}{cc}
\tilde{a}(V) & \tilde{b}(V)\\ \tilde{c}(V) & \tilde{d}(V)
\end{array}\right) =
\left(
\begin{array}{cc}
p_2 & -p_1\\ 0 & p_2^{-1}
\end{array}\right)
\left(
\begin{array}{cc}
a(V) & b(V)\\ c(V) & d(V)
\end{array}\right)
\left(
\begin{array}{cc}
p_2^{-1} & p_1\\ 0 & p_2
\end{array}\right)\label{unit}
\end{equation}
Eq. (\ref{ceq}) states that this transformation can be written as operator (which we will denote $\Omega$) acting in quantum space, i.e.
\begin{equation}
\tilde{b}(V)= \Omega\, b(V) \, \Omega^{-1}\,,
\end{equation}
and $\Omega$ is defined by (\ref{Tsl2}) with
$g = \left(\footnotesize{
\begin{array}{cc}
p_2^{-1} & p_1\\ 0 & p_2
\end{array}}\right)$ in each site. We have
\begin{align}\label{Omega}
\Omega\varphi(x_1,\dots, x_N)=[p_2 - p_1 x_1]^{v_{21} - v_{11} - 1} \dots
[p_2 - p_1 x_N]^{v_{2N} - v_{1N} - 1}
\varphi\left(\textstyle{\frac{p_2^{-1}x_1}{p_2 - p_1 x_1}, \dots,
\frac{p_2^{-1} x_N}{p_2 - p_1 x_N}}\right).
\end{align}
If $\varphi$ is eigenfunction of $b(V)$ then $\Omega\,\varphi$ is eigenfunction of $\tilde{b}(V).$ Eigenfunctions of $SL(2,\mathbb{C})$ operator $b(V)$ were found earlier in sec. 5 so that we reduced the eigenproblem for $B(V)$ to the analogous problem for the algebra with lower rank.

Let us combine together all the parts of the answer.
In last few sections we were constructing eigenfunction for $B(V)$ and now
we should substitute $v_{1k} = u -2-r_k$ and $v_{2k} = u-2-s_k.$   We have
\begin{equation}\label{final}
\Psi_{p_1p_2 b}(q|x, y,z) = W(U, V)\, \Omega\, \varphi(x_1, \dots, x_N),
\end{equation}
where $W(U, V)$ is constructed explicitly, $\Omega$ is
defined by~(\ref{Omega}) and $\varphi$ is eigenfunction of the operator $b(V)$
\begin{align}
\left( \begin{array}{cc} a(V) & b(V) \\ c(V) & d(V) \end{array} \right) =
L_1(\mathbf{v}_1+1)L_N(\mathbf{v}_N+1)\,\cdots\, L_2(\mathbf{v}_2+1)\,,
\end{align}
from the monodromy matrix of the $SL(2, \mathbb{C})$ magnet with matrix of parameters
$$
V = \left( \begin{smallmatrix} v_{11}+1 & v_{1N}+1 & \dots & v_{12}+1 \smallskip\\ v_{21}+1 & v_{2N}+1 & \dots & v_{22}+1 \end{smallmatrix} \right)
$$ and sites ordered as $ 2, \dots, N, 1$ from the right to the left.
If for $\varphi$ eigenvalues are $p, q_i:$
\begin{equation}
b(V)\phi = -ip\prod\limits_{i=1}^{N} (u - q_i)\phi
\end{equation}
then the corresponding eigenvalues of~(\ref{final}) are
\begin{equation}
B(u)\Psi_{p_1 p_2 p}(q|x, y,z) = ip\prod\limits_{i=1}^{N-1}(u - q_i)(u-r_i)(u-s_i)\,   \Psi_{p_1p_2p}(q|x, y,z)
\end{equation}
and
\begin{eqnarray}
E_{31}\Psi_{p_1p_2p}(q|x, y, z) &\!\!=\!\!&-ip_1 \Psi_{p_1p_2p}(q|x, y,z) \nonumber \\
E_{32} \Psi_{p_1p_2p}(q|x, y,z) &\!\!=\!\!& -ip_2 \Psi_{p_1p_2p}(q|x, y, z).\nonumber
\end{eqnarray}

\section{Conclusions}

The main result of the present paper is the construction of the
generalized eigenfunctions of the operator $B(u)$. The system of these eigenfunctions define the kernel of the integral operator, which provides transformation to the representation of separated variables. We have presented the algebraic part of the construction only and the main idea is the following.
Elements of the monodromy matrix
\begin{equation}\nonumber
T(U) = L_N(\mathbf{u}_N)L_{N-1}(\mathbf{u}_{N-1})
\cdots L_{2}(\mathbf{u}_{2}) L_1(\mathbf{u}_1)
\end{equation}
depend on the set of parameters $\mathbf{u}_{i}$, where $1\leq i \leq N$ and we combine all parameters in the matrix $U$.
The Sklyanin B-operator depends on the whole set of parameters
in monodromy matrix $B = B(U)$ but it appears that operators
with different sets of parameters are unitary equivalent
\begin{align}\nonumber
B(U) = W(U, V)\, B(V)\, W^{-1}(U, V)
\end{align}
where operator $B(V)$ depends on the new set of parameters $V$.
Then the generalized eigenfunction of the operator $B(U)$
can be represented in the form
\begin{align}\nonumber
\Psi = W(U, V)\, \Psi_0
\end{align}
where $\Psi_0$ is some particular eigenfunction of the operator $B(V)$.
The matrix $V$ is generic, so that we obtain a sufficiently rich set of eigenfunctions.
The construction of the intertwining operator $W(U, V)$ which
allows to change the set of parameters $U\to V$
\begin{align}\nonumber
B(U)\, W(U, V) = W(U, V)\, B(V)
\end{align}
extensively uses the intertwining operators from
the representation theory of $SL(n,\mathbb{C})$~\cite{GN,KnS,K}.

We have presented only algebraic part of the construction and
there remain many open questions which we hope answer in the future.
Among problems that attracts attention are the following:
\begin{itemize}

\item The investigation of the symmetry of the eigenfunction with respect to permutations of $\{ q_i \}.$

\item Calculation of the proper normalization coefficient for
$\Psi_{p_1 p_2 p}(q| x, y, z)$  that provides
$$
A(q_i)\Psi_{p_1 p_2 p}(q| x, y, z) = \Psi_{p_1 p_2 p}(E^+_i q| x, y, z)
$$
and makes eigenfunctions symmetric with respect to permutations
of $\{ q_i \}.$

\item Proof of the orthogonality of the set of eigenfunctions.
Explicit calculation of the scalar product
which gives the Sklyanin measure.

\item Proof of the completeness of the set of eigenfunctions.

\end{itemize}

In the case of $SL(2,C)$ algebra all such problems except to the completeness
were considered in~\cite{DKM,DM}. The main computational tool for the calculation of scalar products was the Feynman diagram technique.
The calculation of integrals is reduced to the transformation and simplification of the diagrams according some graphic rules (chain integration rule,star-triangle relation). At the moment this Feynman diagrams technique is not worked out in the $SL(3,\mathbb{C})$ case.

A quantum inverse scattering
based method for proving completeness was developed in~\cite{K1,K2}.
We should note that for proving completeness the Mellin-Barnes integral representation~\cite{KharLeb1,KharLeb2,KharLeb3} is well-suited~\cite{K1,K2}
but the Gauss-Givental representation is more useful for proving orthogonality.
In the present paper we have constructed the Gauss-Givental representation for eigenfunctions of the Sklyanin's operator for $SL(3,C)$ magnet. The construction of the Mellin-Barnes integral representation is a separate open problem.

Finally, one should mention the recent works~\cite{SS} where the
unitarity of the b-Whittaker transform is proven and~\cite{DKKM}
where the unitarity of the SOV-transformation for the modular XXZ magnet is proven.

\medskip

\smallskip
{\bf Note added.}

When this paper was written, we learned about the recent work by
J.M. Maillet and G. Niccoli\cite{JMM} in which an alternative
approach to SOV is suggested.
In the paper\cite{JMM} all representations in the
quantum space are finite-dimensional in contrary to our case of
infinite-dimensional principal series representations.
It seems that the detailed investigation of the possible
interrelations will be very instructive.

\medskip

\smallskip
{\bf Acknowledgements.}
This work is supported by the Russian Science Foundation (project no. 14-11-00598).


\begin{thebibliography}{99}


\bibitem{KS}
P. P. Kulish, E. K. Sklyanin, {\it Quantum spectral transform method. Recent developments}, Lect. Notes Phys. {\bf 151} 1982, 61.

\bibitem{Fad1}
L. D. Faddeev, {\it How algebraic Bethe ansatz works for integable model}\\ Quantum symmetries/Symmetries Quantiques, Proc. Les-Houches symmer school, LXIV, Eds. A. Connes, K. Kawedzki, J. Zinn-Justin. North Holland, 1998, 149-211.

\bibitem{KRS}
P.P. Kulish, N.Yu.Reshetikhin and E.K.Sklyanin, {\it Yang-Baxter
equation and representation theory}, Lett.Math.Phys. {\bf 5}
(1981) 393-403.

\bibitem{SKL1}
E. K. Sklyanin, {\it The quantum Toda chain}, \\
Lect. Notes in Phys. 226 (1985), 196–233.

\bibitem{SKL2}
E. K. Sklyanin,{\it Quantum Inverse Scattering Method.Selected
    Topics}, in "Quantum Group and Quantum Integrable Systems" (Nankai
    Lectures in Mathematical Physics), ed. Mo-Lin Ge,Singapore:World
    Scientific,1992,pp.63-97; [arXiv:hep-th/9211111].

\bibitem{SKL3}
E. K. Sklyanin, {\it Separation of variables in the classical integrable SL(3) magnetic chain},
Commun.Math.Phys. 150 (1992) 181-192; [arXiv:hep-th/9211126].

\bibitem{SKL4}
E.~K.~Sklyanin,
{\it Separation of variables in the quantum
integrable models related to the Yangian Y[sl(3)]},
J.\ Math.\ Sci.\  {\bf 80} (1996) 1861
[Zap.\ Nauchn.\ Semin.\  {\bf 205} (1993) 166];
[arXiv:hep-th/9212076].

\bibitem{SKL5}
E. K. Sklyanin, {\it  Separation of variables - new trends},.
Prog.Theor.Phys.Suppl. 118 (1995) 35-60; [arXiv:solv-int/9504001].

\bibitem{Scott}
D. R. D. Scott, {\it Classical functional Bethe ansatz for SL(N): Separation of variables for the
magnetic chain}, J. Math. Phys. 35 (1994) 5831 doi:10.1063/1.530712; [arXiv:hep-th/9403030].

\bibitem{Gekhtman}
M. I. Gekhtman, {\it Separation of variables in the classical SL(N) magnetic chain}, Comm. Math.
Phys. Volume 167, Number 3 (1995), 593-605.

\bibitem{Smirnov}
F. A. Smirnov, {\it Separation of variables for quantum integrable models
related to $ U_q(\hat{sl}_N) $}; [arXiv:math-ph/0109013].

\bibitem{Bgood}
N. Gromov, F. Levkovich-Maslyuk, G. Sizov,
{\it New Construction of Eigenstates and Separation of Variables for SU(N) Quantum Spin Chains},
JHEP 1709 (2017) 111; [arXiv:1610.08032]

\bibitem{Slav}
A. Liashyk, N. A. Slavnov,
{\it On Bethe vectors in $gl_3$-invariant integrable models}, JHEP 1806 (2018) 018; [arXiv:1803.07628]

\bibitem{Maillet}
N. Kitanine, J. M. Maillet, G. Niccoli, V. Terras,
{ \it The open XXX spin chain in the SoV framework: scalar product of separate states},
J. Phys. A: Math. Theor. 50 (2017) 224001
[arXiv:1606.06917].

\bibitem{Nic}
N. Kitanine, J.-M. Maillet, G. Niccoli,
{\it Open spin chains with generic integrable boundaries: Baxter equation and Bethe ansatz completeness from SOV}, J. Stat. Mech., P05015 (2014); [arXiv:1401.4901].

\bibitem{BT}
A.G. Bytsko and J. Teschner,
{\it Quantization of models with non-compact quantum group symmetry. Modular XXZ magnet and lattice sinh-Gordon model}, J. Phys. A 39 (2006), 12927–12982.

\bibitem{KharLeb1}
S. Kharchev and D. Lebedev,
{\it Eigenfunctions of $GL(N, R)$ Toda chain:
The Mellin-Barnes representation},
JETP Lett. 71 (2000), 235–238.

\bibitem{KharLeb2}
S. Kharchev and D. Lebedev,
{\it Integral representations for the eigenfunctions
of quantum open and periodic Toda chains from
QISM formalism}, J.Phys.A 34 (2001), 2247–2258.

\bibitem{KharLeb3}
S. Kharchev, D. Lebedev, and M. Semenov-Tian-Shansky,
{\it Unitary representations of $U_q (sl(2, R))$, the
modular double and the multiparticle q-deformed Toda chains},
Comm. Math. Phys. 225 (2002), 573–609.

\bibitem{Sil}
A.V. Silantyev, {\it Transition function for the Toda chain},
Theor. Math. Phys. 150 (2007), 315–331

\bibitem{KM}
M. Kirch, A.N. Manashov,
{\it Noncompact $SL(2,R)$ spin chain},
JHEP 0406 (2004) 035 ,
e-Print: hep-th/0405030

\bibitem{FK}
L. D. Faddeev and G. P. Korchemsky, {\it High-energy QCD as a completely integrable model},
Phys. Lett. B 342 (1995), 311–322.

\bibitem{L0}
L. N. Lipatov, {\it High-energy asymptotics of multicolor QCD
and exactly solvable lattice models}, Pisma Zh. Eksp. Teor. Fiz. 59 (1994), 571–574 (JETP Lett. 59 (1994), 596–599).

\bibitem{L1} H. J. De Vega and L. N. Lipatov,
{\it Interaction of reggeized gluons in the Baxter-Sklyanin representation}, Phys. Rev. D 64 (2001) 114019.

\bibitem{L2} H. J. de Vega and L. N. Lipatov,
{\it Exact resolution of the Baxter equation for reggeized gluon interactions}, Phys. Rev. D 66 (2002) 074013.

\bibitem{DKM}
S.~E.~Derkachov, G.~P.~Korchemsky and A.~N.~Manashov,
{\it Noncompact Heisenberg spin magnets from high-energy QCD.
I: Baxter Q-operator and separation of variables,}
Nucl.\ phys.\ B {\bf 617} (2001) 375 [arXiv:hep-th/0107193].

\bibitem{DM}
S.E. Derkachov and A.N. Manashov,
{\it Iterative construction of eigenfunctions of
the monodromy matrix for
SL(2,C) magnet}, J.Phys. A: Math. gen. 47 (2014), 305204.
e-Print: arXiv:1401.7477 [math-ph]

\bibitem{GG}
A. Givental,
{\it Stationary phase integrals, quantum Toda lattices,
flag manifolds and the mirror conjecture},
AMS Trans. (2) 180 (1997), 103–115.

\bibitem{VDKU}
P. Valinevich, S.Derkachov, P. Kulish, E. Uvarov,
{\it Construction of eigenfunctions for a system of quantum minors of the monodromy matrix for an  SL(n,C) -invariant spin chain},
Theor.Math.Phys. 189 (2016) no.2, 1529-1553, Teor.Mat.Fiz. 189 (2016) no.2, 149-175

\bibitem{MNO} A.~Molev, M.~Nazarov, G.~Olshanskii, {\it Yangians and classical Lie algebras}, Russ.Math.Surveys 51 (1996) 205

\bibitem{M}
A. Molev, {\it Yangians and Classical Lie Algebras},
Mathematical Surveys and Monographs 143,
AMS, Providence, RI, 2007.

\bibitem{GN}
I. M. Gelfand , M. A. Naimark,
{\it Unitary representations of the classical groups}, Trudy Mat. Inst.
Steklov., vol. 36, Izdat. Nauk SSSR, Moscow - Leningrad, 1950; Gernman transl.: Academie -Verlag, Berlin, 1957.

\bibitem{KnS}
A. Knapp and E. Stein,
{\it Intertwining operators for semi-simple Lie groups},
Ann. of Math. (2) 93
(1971), 489-578.

\bibitem{K}
Knapp A.W. {\it Representation theory of semisimple groups: an overview based on examples}, Princeton,N.J.:Princeton
Univ.Press,1986.

\bibitem{DM1}
S. E. Derkachov and A. N. Manashov,
{\it R-Matrix and Baxter Q-Operators for the Noncompact
SL(N,C) Invarianit Spin Chain}, SIGMA 2 (2006) 084.

\bibitem{DM3}
S. Derkachov and A. Manashov,
{\it General solution of the Yang-Baxter equation with the symmetry
group SL(n, C)}, Algebra i Analiz 21 (4) (2009), 1–94 (St. Petersburg Math. J. 21 (2010), 513–577).

\bibitem{K1}
K.K. Kozlowski,
{\it Asymptotic analysis and quantum integrable models},
e-Print: arXiv:1508.06085 [math-ph]

\bibitem{K2}
K.K. Kozlowski, {\it Unitarity of the SoV transform for the Toda chain},
Comm. Math. Phys. 334 (2015), no. 1, 223–273

\bibitem{SS}
G. Schrader and A. Shapiro,
{\it On b-Whittaker functions},
math-ph:1806.00747.

\bibitem{DKKM}
S. Derkachov, K. Kozlowski, A.Manashov,
{\it On the separation of variables for
the modular XXZ magnet and the lattice Sinh-Gordon models},
e-Print: arXiv:1806.04487 [math-ph]

\bibitem{JMM}
J.M. Maillet, G. Niccoli,
{\it On quantum separation of variables}
e-Print: arXiv:1807.11572 [math-ph]


\end{thebibliography}
\end{document}